\documentclass[fleqn,usenatbib,useAMS]{mnras}

\usepackage{graphicx}	
\usepackage{amsmath}	
\usepackage{amssymb}	
\usepackage{multicol}        
\usepackage{bm}		
\usepackage{pdflscape}	
\usepackage{subcaption}

\usepackage[dvipsnames]{xcolor}

\usepackage[T1]{fontenc}
\usepackage{ae,aecompl}

\usepackage{newtxtext,newtxmath}

\title[Practical Morphology Tools from Deep Learning]{Practical Galaxy Morphology Tools from Deep Supervised Representation Learning}

\author[M. Walmsley]{Mike Walmsley$^{1}$,\thanks{Contact e-mail: \href{mailto:michael.walmsley@manchester.ac.uk}{michael.walmsley@manchester.ac.uk}}
Anna M.~M.~Scaife$^{1,2}$,
Chris Lintott$^{3}$,
Michelle Lochner$^{4,5}$,
\newauthor
Verlon Etsebeth$^{4}$,
Tobias G\'eron$^{3}$,
Hugh Dickinson$^{6}$,
Lucy Fortson$^{7,8}$,
Sandor Kruk$^{9,10}$,
\newauthor
Karen L. Masters$^{11}$,
Kameswara Bharadwaj Mantha$^{7,8}$,
Brooke D. Simmons$^{12}$
\\
$^{1}$Jodrell Bank Centre for Astrophysics, Department of Physics \& Astronomy, University of Manchester, Oxford Road, Manchester M13 9PL, UK\\
$^{2}$The Alan Turing Institute, 96 Euston Road, London NW1 2DB, UK\\
$^{3}$Oxford Astrophysics, Department of Physics, University of Oxford, Denys Wilkinson Building, Keble Road, Oxford, OX1 3RH, UK\\
$^{4}$Department of Physics and Astronomy, University of the Western Cape, Bellville, Cape Town, 7535, South Africa\\
$^{5}$South African Radio Astronomy Observatory (SARAO), The Park, Park Road, Pinelands, Cape Town 7405, South Africa\\
$^{6}$School of Physical Sciences, The Open University, Milton Keynes, MK7 6AA, UK\\
$^{7}$Minnesota Institute for Astrophysics, University of Minnesota, 116 Church St SE, Minneapolis, MN 55455, USA\\
$^{8}$School of Physics and Astronomy, University of Minnesota, 116 Church St SE, Minneapolis, MN 55455, USA\\
$^{9}$Max-Planck-Institut für extraterrestrische Physik, Giessenbachstrasse 1, D-85748 Garching bei München, Germany\\
$^{10}$European Space Agency, ESTEC, Keplerlaan 1, NL-2201 AZ, Noordwijk, the Netherlands\\
$^{11}$Departments of Physics and Astronomy, Haverford College, 370 Lancaster Avenue, Haverford, PA 19041, USA\\
$^{12}$Department of Physics, Lancaster University, Bailrigg, Lancaster, LA1 4YB, UK
}

\date{Last updated 2022 February 21; in original form 2021 October 20}

\pubyear{2021}

\begin{document}
\label{firstpage}
\pagerange{\pageref{firstpage}--\pageref{lastpage}}
\maketitle

\begin{abstract}

Astronomers have typically set out to solve supervised machine learning problems by creating their own representations from scratch.
We show that deep learning models trained to answer every Galaxy Zoo DECaLS question learn meaningful semantic representations of galaxies that are useful for new tasks on which the models were never trained. We exploit these representations to outperform several recent approaches at practical tasks crucial for investigating large galaxy samples.
The first task is identifying galaxies of similar morphology to a query galaxy. 
Given a single galaxy assigned a free text tag by humans (e.g. `\#diffuse'), we can find galaxies matching that tag for most tags.
The second task is identifying the most interesting anomalies to a particular researcher. Our approach is 100\% accurate at identifying the most interesting 100 anomalies (as judged by Galaxy Zoo 2 volunteers).
The third task is adapting a model to solve a new task using only a small number of newly-labelled galaxies. Models fine-tuned from our representation are better able to identify ring galaxies than models fine-tuned from terrestrial images (ImageNet) or trained from scratch. 
We solve each task with very few new labels; either one (for the similarity search) or several hundred (for anomaly detection or fine-tuning). This challenges the longstanding view that deep supervised methods require new large labelled datasets for practical use in astronomy.
To help the community benefit from our pretrained models, we release our fine-tuning code \href{https://github.com/mwalmsley/zoobot}{zoobot}. Zoobot is accessible to researchers with no prior experience in deep learning.

\end{abstract}




\section{Introduction}
\label{sec:introduction}

The core of many machine learning approaches is learning to calculate useful representations, i.e. lower-dimensional summaries of images or other data with which a prediction can be made.
Learning hierarchical representations, where the representation learned by one layer becomes the input to the next, is the cornerstone of deep learning.
Representations are particularly important for words and images, where the input feature space is high-dimensional and thus more difficult to make direct predictions with than, for example, typical tabular data \citep{LeCun2015, GoodfellowBook2016}.

To date, astronomers have typically set out to solve supervised machine learning problems by creating their own representations from scratch.
They often train a randomly-initialised model only on the labelled data they are directly interested in.
This is often true even for researchers solving similar tasks with similar methods on similar datasets. For example, distinguishing between early and late-type galaxies in SDSS imaging \citep{Khalifa2018, Sanchez2018, Fischer2018, Khrmatsov2019, Variawa2020, Barchi2020, Walmsley2020}.
The expressivity of each model is limited by the size of the training data (to prevent overfitting) which in turn limits performance on complex tasks requiring such expressivity. 
We aim to demonstrate in this paper that, under certain conditions, starting from a representation learned elsewhere is more effective; specifically, that  exploiting representations learned while solving a broad set of galaxy morphology tasks can dramatically improve performance on new morphology tasks.

We are primarily motivated by results from the natural language community.
Recent empirical research suggests that the performance of deep natural language models with Transformer architectures follows fundamental scaling relations \citep{Vaswani2017}.
Broadly speaking, performance increases approximately as a power law with respect to the number of model parameters, the size of the training dataset, and the computational budget \citep{Kaplan2020}.
For example, increasing the number of model parameters will likely increase performance provided one has access to effectively unlimited data and compute.
Most researchers have neither, and so the best-performing models are increasingly created by a few well-resourced groups such as OpenAI \citep{Brown2020} and Google Brain \citep{Fedus2021}.
These natural language models are trained to predict masked words in sentences (along with related tasks) and so effectively all digitised writing is useful training data. This style of training is known as `self-supervised' as the model is trained in a supervised manner on labels (masked words) already present in the data itself.
Having learned an effective representation of language, the models can then be fine-tuned, i.e. gradually adapted with additional data, on so-called domain tasks: tasks of practical interest such as summarising news articles, coding websites, or understanding emotion \citep{Kant2018,YangZ2020,Austin2021}.
Crucially, because the fundamental language representation is already learned, fine-tuning achieves state-of-the-art performance using far more modest data and compute than training from scratch. 

Could such an approach work for galaxy morphology? 
Can we train models on large datasets of galaxy images and then use the learned representations as a starting point to solve new practical morphology tasks?
Convolutional neural networks (CNNs), the now-standard approach for classifying galaxy images, likely follow similar scaling laws \citep{Sharma2020}.
It is possible to train a CNN on images in an analogous self-supervised manner by predicting pixel values (generative learning, e.g. \citealt{VanDenOord2016a}) or by enforcing that randomly-transformed images retain similar representations (contrastive learning, e.g. \citealt{Chen2020}). In astronomy, this is typically done in the context of solving a particular task \citep{Zanisi2021,Sarmiento2021}, though recent work by \cite{Hayat2021self} uses self-supervised learning to learn galaxy representations explicitly for generic downstream tasks.

One important drawback to self-supervised methods, and unsupervised methods more broadly, is that the representations must be learned directly from image pixel values and so it is difficult to create representations informed by our physical understanding of the world.
We believe this may lead to predictions that do not make physical sense.
For example, \cite{Buncher2020}, aiming to predict how a shallow galaxy image would appear in a deeper survey, found their unsupervised generative model would fill in large artefacts in the original images with a plausible sky background.
\cite{Spindler2020} found their unsupervised generative model clustered galaxies according to whether they have a background partner galaxy in the top or bottom corner of the image.
Contrastive learning allows a degree of physics input through the choice of augmentations, but these are typically limited to basic invariances (e.g. flips, rotations, added noise, etc).
We would prefer a representation informed by our human understanding of an image, beyond the raw pixels themselves: a representation that `understands' that a background partner galaxy is not scientifically relevant.

Supervised methods present an alternative way to learn representations. Their representations are optimised for the supervised task and so are more strongly influenced by human labels, which are designed to focus learning on the most scientifically relevant aspects (e.g. bars, arms, etc.).
One is then faced with the apparent dilemma of learning a representation using either self-supervised approaches with near-limitless data but limited physical understanding, or supervised approaches with less data but scientifically relevant labels.

To minimise the number of labels required to learn meaningful representations from supervised approaches, one can exploit existing labelled datasets.
Pretraining on ImageNet \citep{Russakovsky2015}, a relatively\footnote{As compared to other benchmark datasets e.g. MNIST, CIFAR10. Diversity in ImageNet is the subject of significant attention (e.g. \citealt{Recht2019, YangFair2020}), in part because of its widespread use.} diverse benchmark dataset containing images of 1000 terrestrial classes, is particularly common in the computer science literature \citep{Marmanis2016,Tschandl2019,Mathis2020,Ridnik2021}. 
Astronomers have recently experimented with pretraining on ImageNet to better solve astronomical tasks.
\cite{Ackermann2018} and \cite{Martinazzo2020} each measured the performance on galaxy-morphology-related tasks of CNNs either initialised randomly or pretrained on ImageNet. In both cases, the ImageNet-pretrained CNN performed significantly better. Additionally, \cite{Wu2018} noted that using frozen ImageNet weights for the first four layers of their CNN improved performance for cross-matching sources given a fixed amount of training time\footnote{This does not necessarily imply improved performance at convergence, however, as pretrained models may converge faster \citep{HeRethinking2019}.}.
Astronomers have also found success with pretraining CNNs on a previously-labelled survey and used them to solve the same task for a new survey:
\cite{Sanchez2019} pretrained on SDSS and fine-tuned to DES,
\cite{Perez-Carrasco2018} pretrained on CANDELS and fine-tuned to CLASH, and
\cite{Tang2019} pretrained on NVSS and fine-tuned to FIRST (and vice versa).

We hypothesise that ImageNet pretraining works well for new terrestrial tasks because the classification task is broad (i.e. distinguish 1000 classes including `toilet paper' and `triceratops') and so the representation is likely to be appropriate for new terrestrial classes, but will work less well for galaxy morphology tasks because the terrestrial-trained representation is less appropriate. ImageNet classes have dramatically different shapes, textures and signal-to-noise levels than galaxies, and so only the most basic representations (edges, curves, etc., detected by the first convolutional layers, \citealt{HeRethinking2019}) are likely to be useful.
On the other hand, we believe that while pretraining on other galaxy surveys is helpful because the representations learned will be appropriate for galaxy images, the classification task is narrow (e.g. distinguish mergers from non-mergers) and so the representations are likely to be specific to that task.  

\textbf{The core argument of this paper is that general purpose supervised galaxy morphology representations would be better learned from solving a broad galaxy morphology task}. These representations would be both more relevant to galaxy images than are learned from ImageNet (which is comprised of terrestrial images) and more widely applicable to new morphology tasks than representations learned from a single narrow morphology task (as in previous work). 

We argue that the models trained by \cite{Walmsley2022decals} on Galaxy Zoo DECaLS have learned to answer just such a broad task and thus provide ideal cross-task representations. The Galaxy Zoo DECaLS project asked volunteers on \href{www.galaxyzoo.org}{www.galaxyzoo.org} a diverse set of questions designed to capture the essential phenomenological features of galaxy morphology such as bars (strong and weak), spiral arms (counts, winding), bulge size, inclination, and so forth. Models were then trained to answer these diverse questions. To do so, the models learned to create general representations suitable for morphology tasks beyond the questions themselves, just as ImageNet classifiers learn general representations for terrestrial tasks beyond identifying the ImageNet classes.

We first investigate the DECaLS models' representations and show that visually similar galaxies are mapped to similar parts of feature space, even for morphology aspects not explicitly measured by the Galaxy Zoo questions (Section \ref{sec:representation_evidence}). We then go on to use that representation to develop and demonstrate practical scientific tools for similarity searches (Section \ref{sec:similarity}), anomaly detection (Section \ref{sec:anomaly}), and transfer learning (Section \ref{sec:transfer}). We share our code and data in Section \ref{sec:data_availability}.

\section{Representations and Visual Similarity}
\label{sec:representation_evidence}

Image representations are crucial for many practical tasks of interest to astronomers.
An image representation function maps the information content of a high-dimensional image to a lower-dimensional vector. A useful representation should allow for the definition of a meaningful distance metric, i.e. similar images should be closer in representation space than dissimilar images, and small changes to an image should lead to small changes in the representation and vice versa. 

In this section, we present evidence that the GZ DECaLS models trained in \cite{Walmsley2022decals} (hereafter W+22) learn such a representation for galaxies. We then use that representation to introduce a method for identifying objects that are similar to a user-selected query galaxy, and demonstrate the method's effectiveness on a diverse and independently-selected set of galaxies.

\subsection{Data}
\label{sec:decals_data}

Throughout this paper, we experiment with galaxies sourced from the Dark Energy Camera Legacy Survey (DECaLS) DR5 \citep{Dey2018}. The selection and image acquisition process is described in detail in W+22. Briefly, galaxies are selected from the NASA-Sloan Atlas v1.0.1 \citep{Albareti2017} if they have an angular radius of \texttt{petrotheta} $> 3$ arcsec and have been observed by DECaLS in the $grz$ bands as of DR5. FITS images are downloaded from the DECaLS cutout service at native telescope resolution with the visible sky area set according to galaxy angular radius\footnote{$s = \max(\min(p_{50} * 0.04, p_{90}*0.02), 0.1$), where \texttt{petro50} and \texttt{petro90} are the NSA columns measuring 50\% and 90\% Petrosian radii}, interpolated to 424x424 pixel thumbnails, and finally rescaled and colourised for human viewing on Galaxy Zoo. 

Unlike GZ DECaLS, we apply an $r$-band magnitude cut of $14.0 < r < 17.77$. 
The fainter limit ensures galaxies are within the bulk of the population with SDSS spectroscopy \citep{Albareti2017} and the brighter limit excludes galaxies with unreliable radii measurements and fields saturated by nearby stars. We also exclude galaxies flagged as likely to be incorrectly sized due to photometric errors in the NASA-Sloan Atlas (see W+22). The resulting catalogue includes 305,657 galaxy images.

For our anomaly detection algorithm (Sec.~\ref{sec:anomaly}), to directly compare our performance with that of \texttt{Astronomaly} \citep{Lochner2021}, we also experiment with the 60,000 Galaxy Zoo 2 images shared as a public training set for the Kaggle `Galaxy Challenge' competition\footnote{https://www.kaggle.com/c/galaxy-zoo-the-galaxy-challenge/data}. The construction and selection of these images is described in \cite{Willett2013} and \cite{Dieleman2015}, respectively. 

\subsection{Calculating Representations}
\label{sec:feature_predictions}

The trained GZ DECaLS models must internally represent galaxies in a way that is appropriate for predicting the answers to GZ DECaLS questions. Here, we describe how we extract those representations. The procedure is essentially identical to making predictions, except that we save the activation values before the final layer rather than the predictions themselves.

Galaxy images are passed to the model following the same procedure with which the model was trained, described in detail in W+22. Briefly, images are converted to greyscale, resampled from 424 to 300 pixels on a side, and then cropped about a random centroid to 224 pixels across (effectively zooming the image by 25\%). This provides an image with an appropriate field-of-view and with input dimensions matching those for which our chosen model architecture was designed. Each time an image is loaded into memory, it is uniquely augmented with an aliased rotation through a random angle and randomly-selected horizontal and vertical flips.

W+22's models use the EfficientNetB0 architecture \citep{Tan2019a}. EfficientNet is composed of a series of mobile inverted bottleneck blocks \citep{Sandler2018}, comparable to standard convolution \& pooling blocks. These stacked blocks are followed by a 1x1 convolutional layer \citep{Szegedy2015Deeper} with 1280 filters, the output of which is global average-pooled (i.e. each filter is replaced with the mean of that filter's activations) for a 1280-dimensional vector. \textbf{This vector is what we refer to throughout as the learned representation}. In normal use, this vector would form the input for the final dense layer, and the outputs of that dense layer would be interpreted as the model predictions. Here, however, we remove the final dense layer and directly record the 1280-dimensional internal representation.

Unlike W+22, which was concerned with predicting well-calibrated posteriors for galaxy morphologies, we do not use dropout or model ensembling to marginalise over the network weights. Instead, we use a single forward pass from a single model. Marginalising might in principle improve performance by removing feature-space noise from the specific weights and augmentations used, but the effect of averaging representations is unclear and so we defer this to future work. We use the weights of a GZ DECaLS model trained on all labelled galaxies (i.e. both training and validation sets) and used by W+22 as part of the ensemble for creating the GZ DECaLS automated catalogue. We refer to this model as `the DECaLS model' in this work.

In this section (for similarity searches) and the following section (Sec. \ref{sec:anomaly}, for anomaly detection), we treat the representation as fixed and therefore precalculate and store the representation for each galaxy. We do not need to make any further CNN predictions when our methods are applied, removing the significant time and hardware requirements typically associated with applying CNN. In Section \ref{sec:transfer} we investigate fine-tuning the representation for improved performance.

\subsection{Visualising}
\label{sec:visualizing}

We use the dimensionality reduction algorithm \texttt{umap} \citep{McInnes2018} to visualise the representations learned by the DECaLS model. \texttt{umap} attempts to balance local and global structure (i.e. distances to close neighbours vs. far neighbours) when compressing a higher-dimensional space. \texttt{umap} is commonly used for visualising high-dimensional spaces in both computer science and astronomy (e.g. \citealt{Clarke2020, Reis2019}).

We assume that the 1280-dimensional ($\mathcal{D}=1280$) representation includes some redundant information because we imposed no independence requirements or weight decay during network training. 
We therefore first compress the representation space with incremental principal component analysis \citep{Ross2008} to $\mathcal{D}=15$ while preserving 98\% of the initial variation. We find this gives more compelling visualizations than using \texttt{umap} directly.

\begin{figure*}
    \centering
    \includegraphics[width=\textwidth]{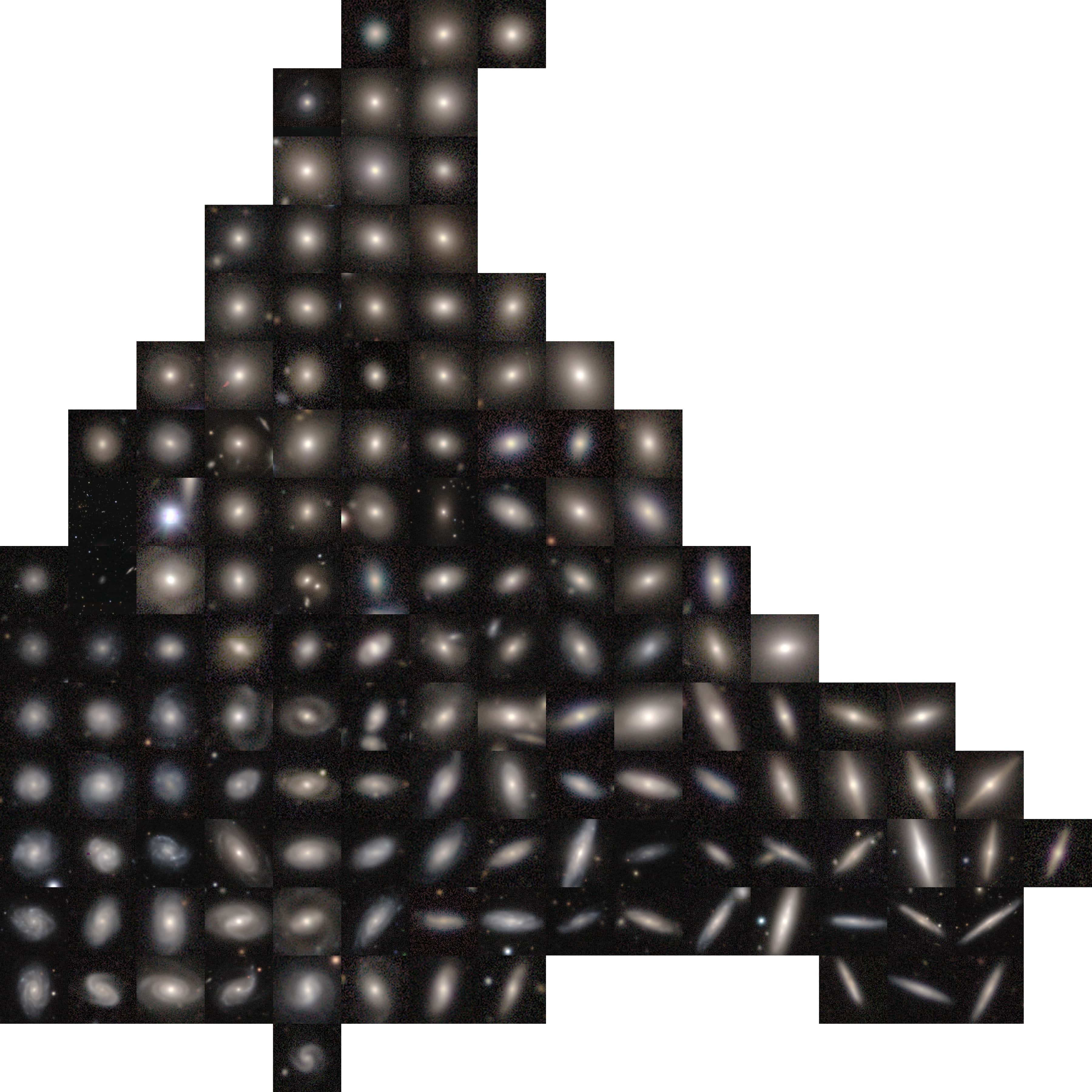}
    \caption{Visualisation of the representation learned by our CNN, showing similar galaxies occupying similar regions of feature space. Created using Incremental PCA and \texttt{umap} to compress the representation to 2D, and then placing galaxy thumbnails at the 2D location of the corresponding galaxy. }
    \label{fig:umap_viz_all}
\end{figure*}

Having compressed the representation from $\mathcal{D}=1280$ to $\mathcal{D}=15$ with incremental PCA and then to $\mathcal{D}=2$ with \texttt{umap}, we can inspect how the representation corresponds to visual appearance by showing galaxy thumbnails located according to their position in the compressed representation.
Figure \ref{fig:umap_viz_all} shows the result for all galaxies. The effect of visual appearance is clear: smooth ellipticals occupy the upper corner, flocculent spirals occupy the lower left, rings and diffuse disks occupy the lower centre, and edge-on-disks occupy the right corner. Figures \ref{fig:umap_viz_feat} and \ref{fig:umap_viz_spiral} show equivalent plots for galaxies filtered (using the GZ DECaLS automated vote fractions from W+22) to include only featured or spiral galaxies, respectively, and show similarly striking visual arrangements within each class.

We conclude that even after compression from $\mathcal{D}=1280$ to $\mathcal{D}=2$, visual similarity strongly affects location in representation space. In the following section we exploit this representation to identify visually similar galaxies.

\subsection{Similarity Searches}
\label{sec:similarity}

Automatically quantifying the similarity of two galaxies is a longstanding but elusive goal.
The most obvious use for quantified similarity is searching for counterparts to known rare objects.
The serendipitous discovery of qualitatively new sources such as Hanny's Voorwerp \citep{Lintott2009} raises the inevitable question `are there more?'
Effective searches for the most similar galaxies \citep{Ardizzone1996,Csillaghy2000,AbdElAziz2017} allow us to make the leap from a one-off curiosity to a new class of objects.
Quantifying similarity is also foundational to any effort at creating automated clusters or taxonomies of galaxies, a topic of much recent interest \citep{Schutter2015,Hocking2017AnLearning,Ralph2019,Cheng2020a, Spindler2020}. The hope is that automated analysis of large-scale modern surveys will reveal galaxy populations that are more objective, and perhaps better connected to the underlying physics of galaxy formation, than the Hubble sequence and its extensions.

What makes two galaxies similar?
Physically meaningful similarity implies not just similar pixels but similar morphology.
Our representations provide a new opportunity for measuring morphological similarity.
Since galaxies of similar morphology have similar representations, we can use the distance in representation space as an estimate of similarity.
We can therefore retrieve the most similar galaxies to a given galaxy simply by listing its nearest neighbours in representation space.

Identifying nearest neighbours in the $\mathcal{D}=1280$ CNN representation is computationally expensive, even with efficient algorithms like \texttt{sklearn}'s KDTree. For convenience, we reduce the dimensionality using Incremental PCA (as we did prior to applying \texttt{umap} in Sec. \ref{sec:visualizing} above). Any choice of PCA dimensionality above $\mathcal{D} \geq 10$ (84\% variance preserved) has a minimal effect on the 50 closest neighbours, while reducing the dimensions from $\mathcal{D}=1280$ to $\mathcal{D}=\mathcal{O}(10)$ reduces the time per search from $\mathcal{O}$(hours)\footnote{Calculating the 50 closest neighbours takes approximately one hour on a standard laptop: fast enough to be possible, but slow enough to be inconvenient.} to $\mathcal{O}$(seconds). We use $\mathcal{D} = 10$ here.

We choose the Manhattan distance $\sum_i | p_i - q_i |$ as our distance metric, implying that similarity is linearly proportional to distance. The Manhattan distance is theoretically preferable to the Euclidean distance for nearest-neighbour searches in high dimensions \citep{Aggarwal2001}. We also experimented with the Euclidean distance and could not confidently identify a qualitative difference in the similarity of the galaxies returned using each distance metric.

The Galaxy Zoo Talk forum \footnote{https://www.zooniverse.org/projects/zookeeper/galaxy-zoo/talk/} provides an independent and diverse selection of galaxies with which to test our similarity search. When writing forum posts about galaxies, Galaxy Zoo volunteers can choose to use `tag' phrases prefaced with a hash, e.g. `\#starforming', analogously to Twitter hashtags. 
For each of the most commonly-used tags (`\#starforming', `\#disturbed', etc.), we use the galaxy most commonly given that tag as our query galaxy and search for similar galaxies in our compressed $\mathcal{D}=10$ representation space. 

The results from those similarity searches are shown in Figure~\ref{fig:similar_tags}. We successfully find similar galaxies in almost all cases. This includes cases like `\#dustlane' where the feature in question is highly specific; most of the returned galaxies are not just edge-on-spirals but edge-on-spirals with dust lanes.

These searches are representative of the typical performance of our method. We do not `cherry-pick' the searches with the best outcomes. We only exclude tags for being related to data not in the image (`\#agn', `\#decals', etc.) or being directly equivalent to a decision tree question (`\#spiral', `\#merging', etc.). Tags have also been grouped semantically (e.g. `\#dust-lane' to `\#dustlane', `\#ringed' to `\#ring', etc.). Figure~\ref{fig:similar_tags} otherwise simply shows searches for the most popular 18 tags.

We emphasise that the DECaLS model \emph{was not explicitly trained on any of these tags}. The model was only trained to predict volunteer votes to the (different) questions in the GZ decision tree (see W+22). In these similarity searches, the model is identifying similar objects based on only a single example: the query galaxy itself. In computer science terminology, the model is performing one-shot learning on a fixed embedding \citep{Feifei2006}.

The occasional failures help us understand what the model can and cannot recognise, which speaks to model interpretability. For example, with the `\#overlapping' example, the volunteers are likely referring to the small companion galaxy (centre-left) but the search returns additional irregular galaxies rather than additional galaxies with small companions. The similarity search is more successful at `\#diffuse' where the query image includes a pair of substantively-sized galaxies and the search returns similar interacting pairs. We can infer that the DECaLS model likely focuses on the main galaxies in the image and has a lower limit for how small (in angular size) a background galaxy can be to affect the representation.

The `\#asteroids' example, where volunteers selected the image due to the small colourful speckle, also illustrates this effect. 
Further, the model is only provided greyscale images, and so could not identify similar colourful speckles even without the size issue above. One could address the lack of sensitivity to colour by training on colour images, at the cost of potential bias for users who prefer a colour-insensitive similarity search (e.g. for investigating links between morphology and star formation).

\begin{figure*}

    \centering
    \includegraphics[width=0.8\textwidth]{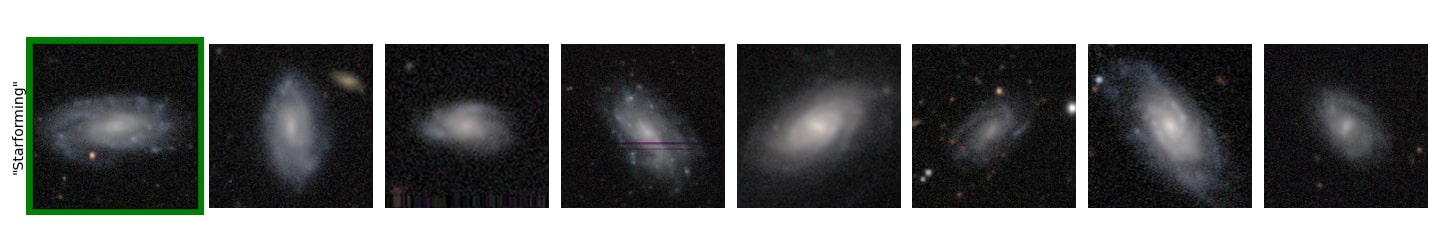}
    \includegraphics[width=0.8\textwidth]{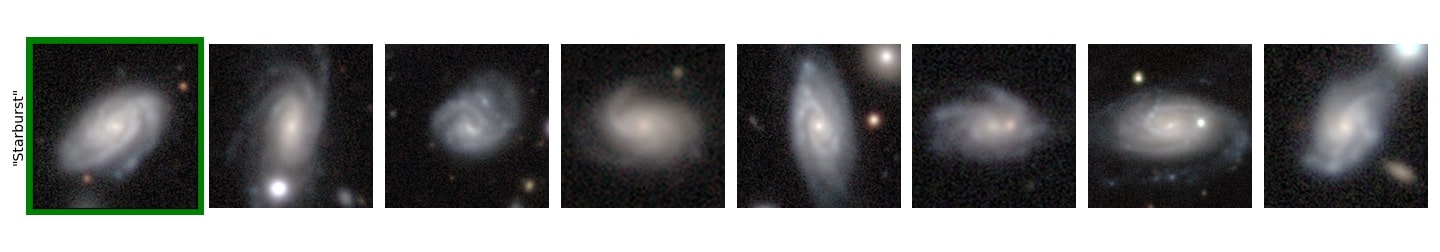}
    \includegraphics[width=0.8\textwidth]{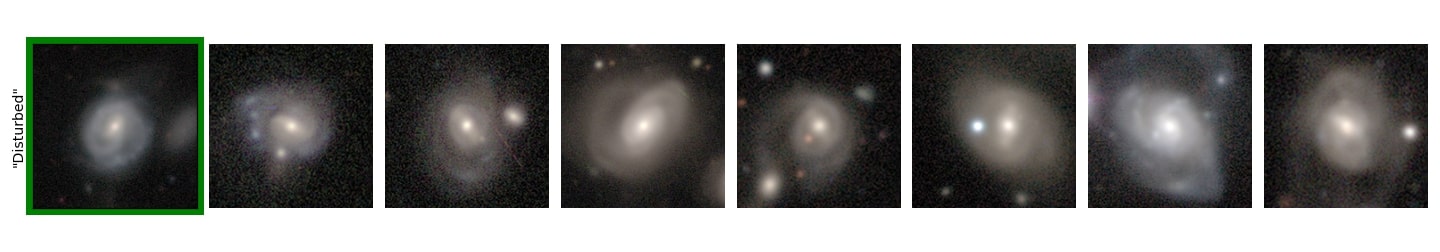}
    \includegraphics[width=0.8\textwidth]{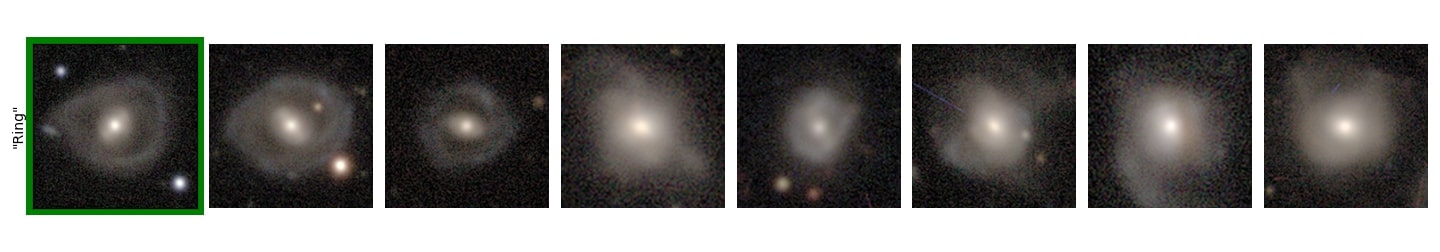}
    \includegraphics[width=0.8\textwidth]{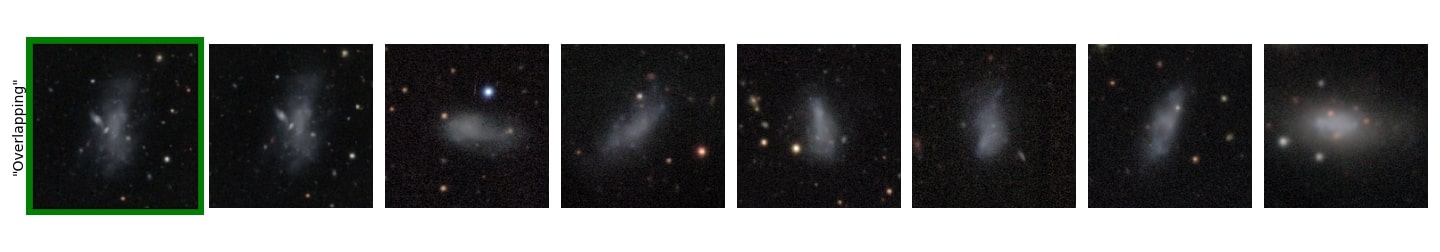}
    \includegraphics[width=0.8\textwidth]{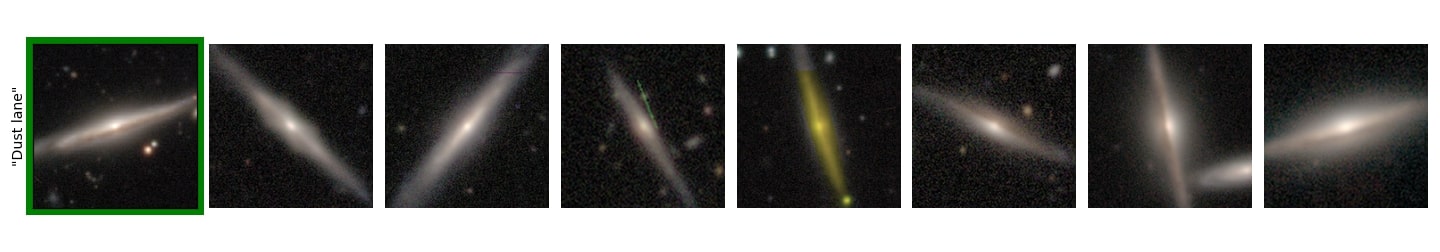}
    \includegraphics[width=0.8\textwidth]{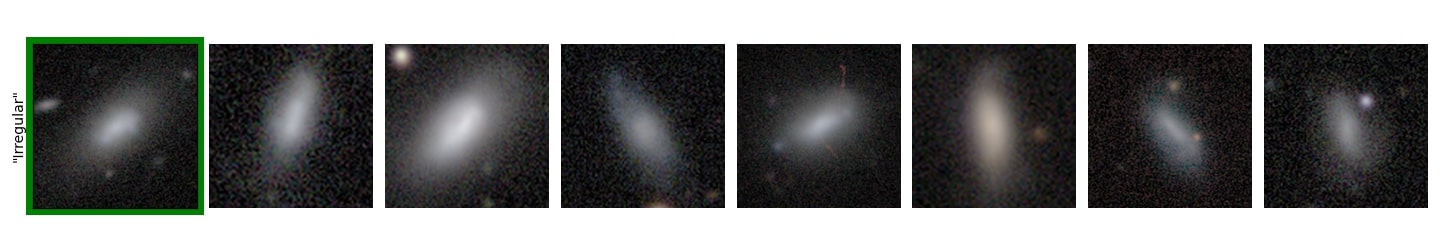}
    \includegraphics[width=0.8\textwidth]{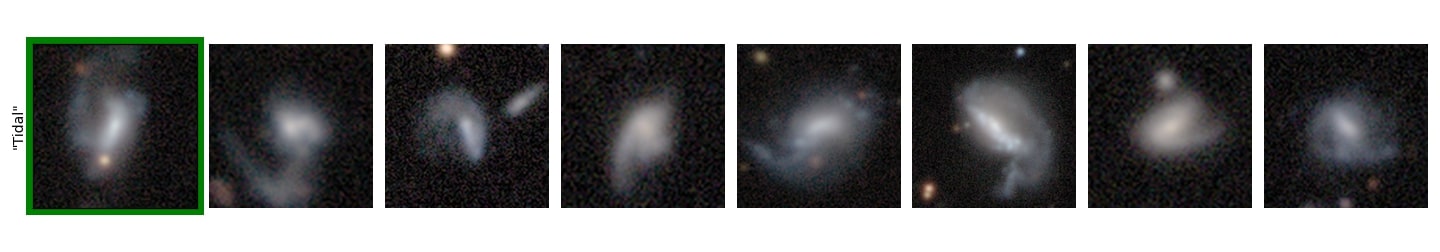}
    \includegraphics[width=0.8\textwidth]{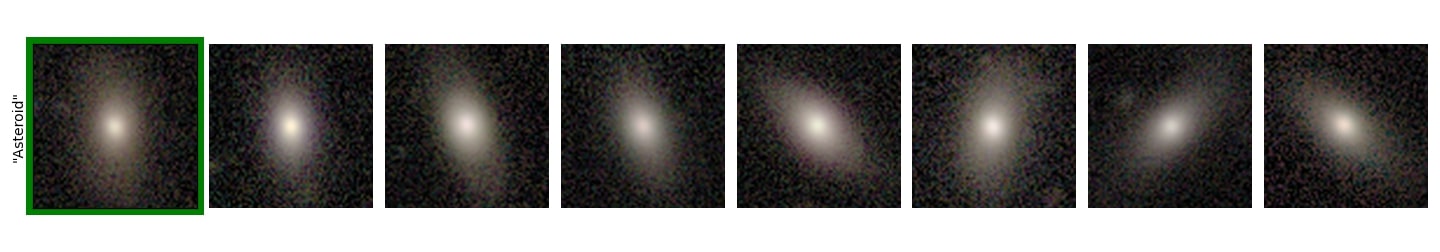}
    
    \caption{Similarity search results for the most common volunteer tags (on which the model was not trained). The query galaxy (left, green border) is the galaxy for which the volunteers most used that tag). The other galaxies on each row are those expected by the DECaLS CNN to be most similar i.e. with the least separation to the query galaxy in representation space. The repeated `overlapping' galaxy is not an error; the background and foreground galaxies are both independently listed in the catalog and identified as similar.}
    \label{fig:similar_tags}
    
\end{figure*}

\begin{figure*}
    \ContinuedFloat
    \centering
    
    \includegraphics[width=0.8\textwidth]{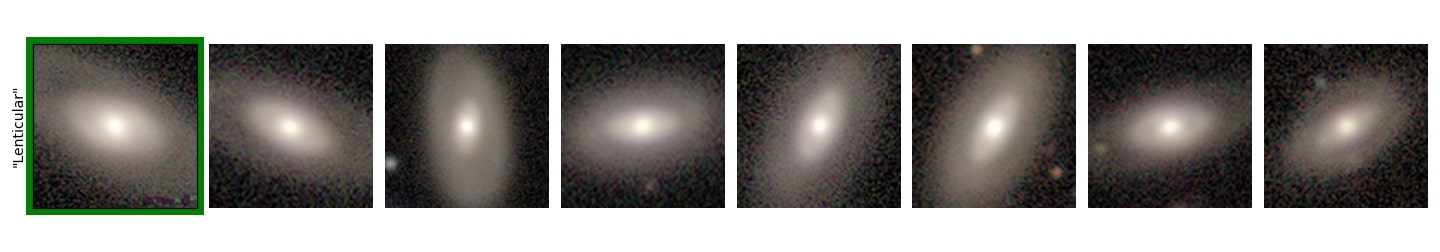}
    \includegraphics[width=0.8\textwidth]{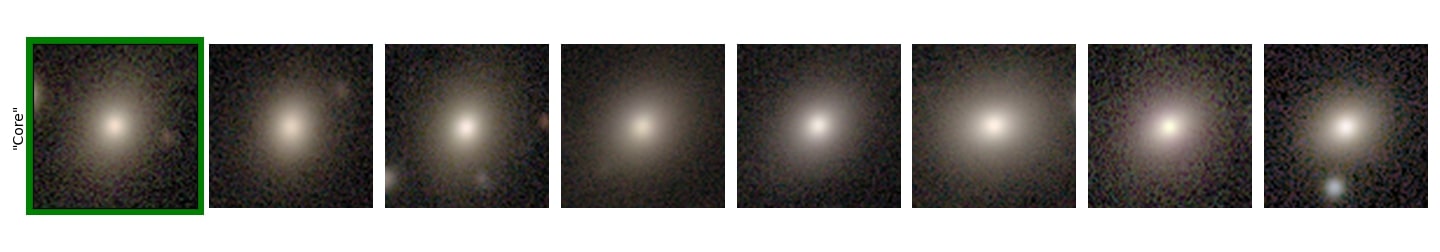}
    \includegraphics[width=0.8\textwidth]{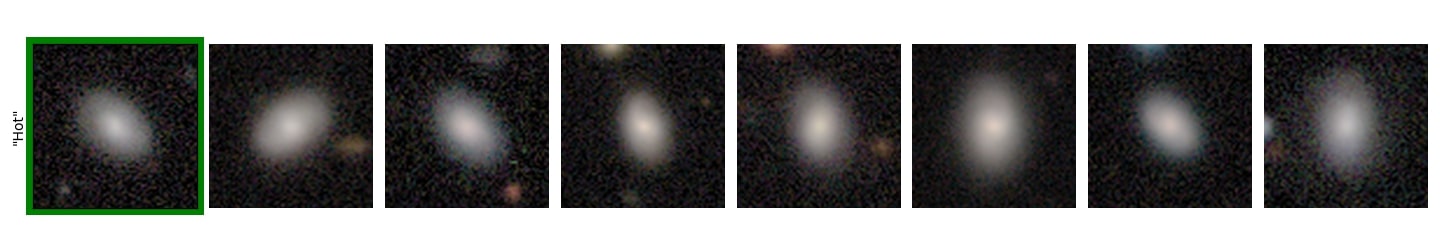}
    \includegraphics[width=0.8\textwidth]{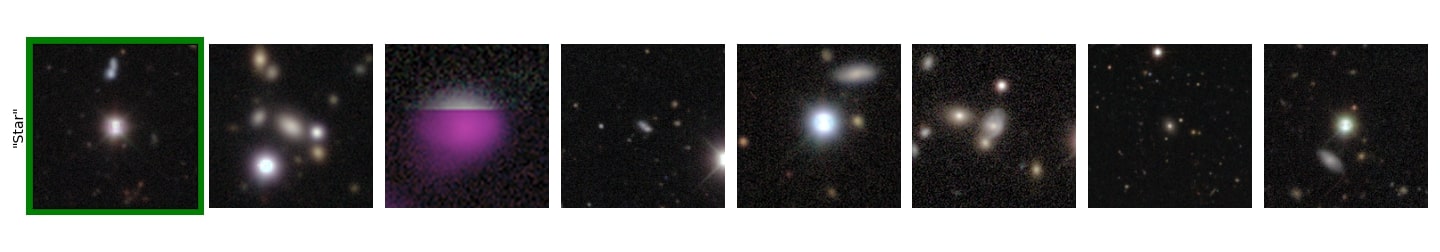}
    \includegraphics[width=0.8\textwidth]{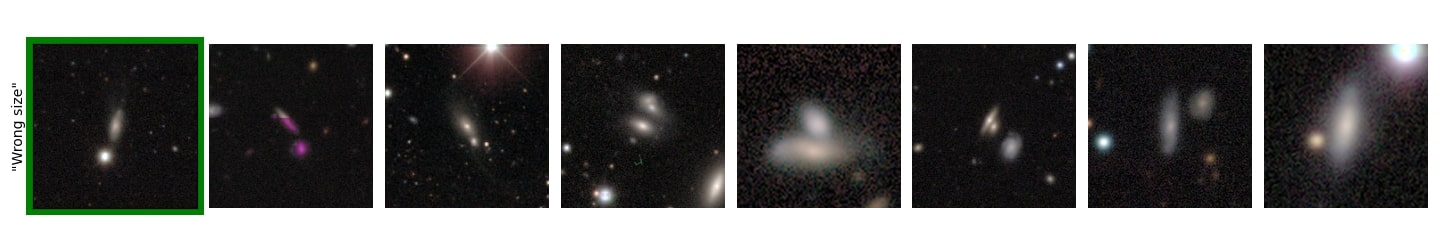}
    \includegraphics[width=0.8\textwidth]{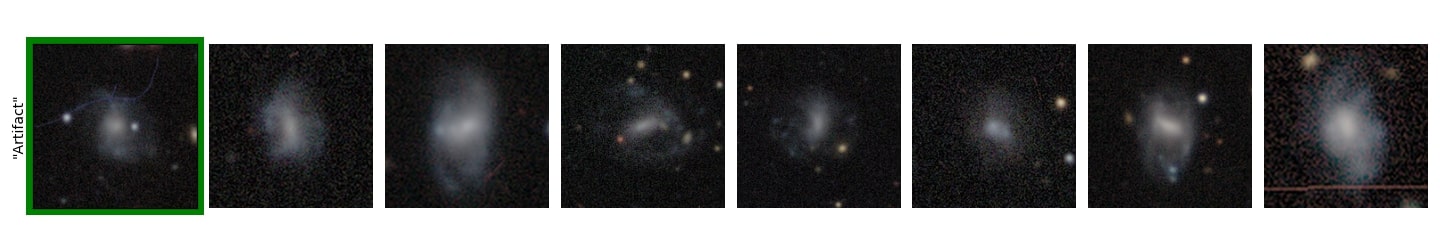}
    \includegraphics[width=0.8\textwidth]{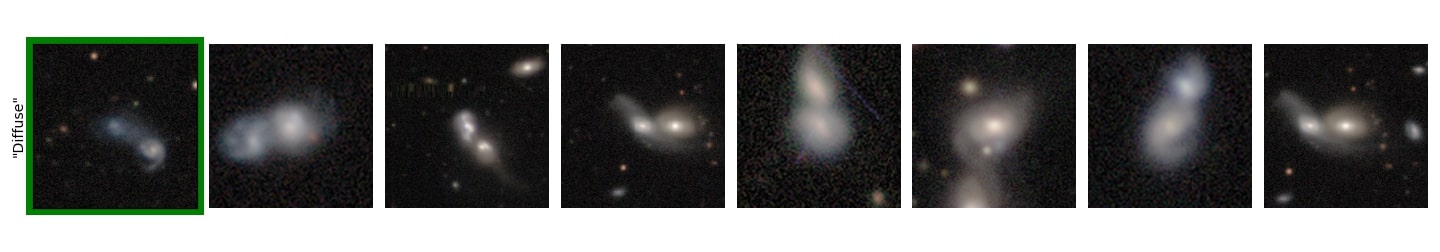}
    \includegraphics[width=0.8\textwidth]{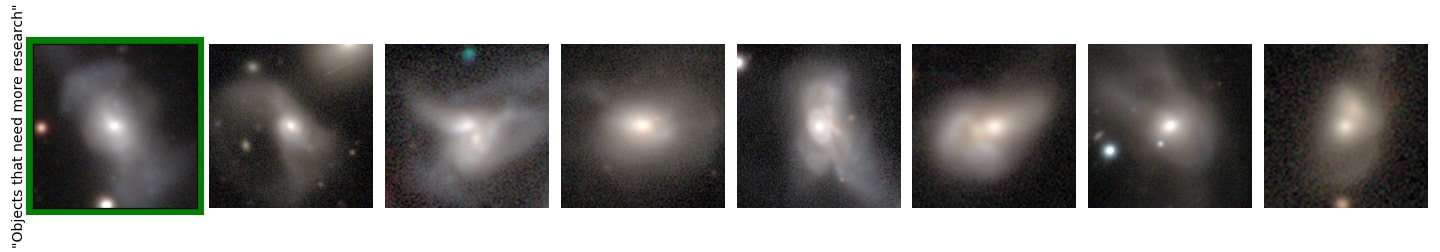}
    \includegraphics[width=0.8\textwidth]{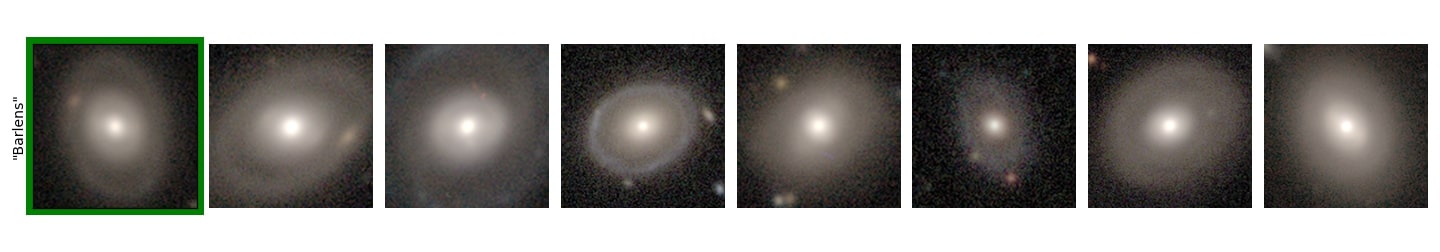}

    \caption{\textbf{(continued)} Similarity search results for the most common volunteer tags (on which the model was not trained). The query galaxy (left, green border) is the galaxy for which the volunteers most used that tag). The other galaxies on each row are those expected by the DECaLS CNN to be most similar i.e. with the least separation to the query galaxy in representation space.}
\end{figure*}

We provide a public interface to our similarity search at \url{https://share.streamlit.io/mwalmsley/decals_similarity/main/similarity.py}. Users enter the coordinates of their desired query galaxy which are then matched to the closest-on-sky DECaLS galaxy in our sample. Images and a table of the most similar galaxies are then returned. Code and instructions for a self-hosted version are available at \url{https://github.com/mwalmsley/decals_similarity}.

\section{Finding Interesting Anomalies}
\label{sec:anomaly}

\subsection{Context}

We showed in Sec. \ref{sec:similarity} that if we have a single example galaxy, we can find similar examples. But what if we don't know what we are looking for?

Many fundamental insights have been driven by rare objects found serendipitously in `big data' catalogs \citep{Cardamone2009,Welsh2011,Boyajian2016}. Such searches may be assisted by machine learning methods aimed at identifying rare objects \citep[e.g.][]{Henrion2013, Baron2017, Storey-Fisher2021AnomalyNetworks}. However, not all rare objects are useful. Instrumental artifacts, completely smooth ellipticals and highly disturbed post-mergers are all `anomalous' in the technical sense of deviating from the typical distribution of images\footnote{Each of these classes were routinely identified as anomalies by common anomaly-finding approaches during the development of this section.}. Only some of these will be valuable to the user. It is therefore crucial that any automated search for anomalies takes into account the user's interests. 

Finding interesting anomalies guided by user feedback is the focus of the computer science subfield of active anomaly detection \citep{Kong2020}. Previous work has considered various schemes where traditional unsupervised anomaly detectors (e.g. Isolation Forests, \citealt{Liu2008}) are coupled to user feedback in an \emph{active learning loop} where the algorithm identifies rare datapoints, those rare datapoints are rated for interest, and a supervised algorithm is trained based on those ratings \citep{Pelleg2014, Das2016, Das2017, Siddiqui2018}. 
The recently-introduced software package \texttt{Astronomaly} (\citealt{Lochner2021}, hereinafter LB21) applies this approach in an astronomical context.

\texttt{Astronomaly} has two parts: the first is a browser interface where users can express their interest in images or one-dimensional data; the second is a set of data processing components which can be configured to extract features, identify rare datapoints, and model user interest. Together, they can be used to apply the active learning loop described above to galaxy images, spectra, lightcurves etc.

\texttt{Astronomaly} is intended as a general anomaly-finding framework which astronomers can extend to suit their specific science goals. Here, we show how several extensions motivated by our new galaxy representations make an \texttt{Astronomaly}-style approach significantly better at identifying merging galaxies, as measured on the benchmark task introduced by LB21. We then show how our improved approach can also find mergers, rings and irregular galaxies in the DECaLS survey. 

\subsection{Method}

In this section, we develop a method to search our CNN representation for anomalies likely to be interesting to a specific user. We build a model of interest as a function of representation by intelligently asking that user about their interest in the galaxies which best help narrow down their preferences (active learning). We then predict their interest in every galaxy to estimate which galaxies they most care about. \\

We will contrast our method with the specific method used by LB21 to demonstrate the quantitative performance of \texttt{Astronomaly} on Galaxy Zoo 2 data, which we will simply call `Baseline'. We describe the task itself and compare results in Sec. \ref{sec:astronomaly_gz2}. Baseline is a particular choice of \texttt{Astronomaly} components designed to work well with this galaxy morphology task while being simple and applicable to other images. 

The general approach of \texttt{Astronomaly} (as in LB21) is as follows.
Galaxy images are converted to features and ranked by rarity. The rarest galaxies are rated by the user according to personal interest. A regression model is fit to these rare galaxies of known interest to predict interest for all other galaxies. Finally, the predicted user interest is combined with the machine learning rarity scores to find galaxies with both high expected interest and high rarity i.e. interesting anomalies.
Those top galaxies themselves can then be rated to continue the active learning cycle of labelling, estimating interest, and choosing new galaxies to label.

LB21's specific Baseline approach chose ellipse fitting as a feature extractor, an Isolation Forest to rank by rarity, and a Random Forest \citep{Breiman2001} to model user interest. We replace each of these steps. We also qualitatively change \texttt{Astronomaly}'s novel active learning approach from labelling the galaxies thought to be most interesting to labelling the galaxies \emph{which, if labelled, would best help to find those interesting galaxies}.\\

\texttt{Astronomaly}'s ellipse-fitting feature extractor works, in short, by placing a series of ellipses enclosing increasing proportions of flux, and recording the properties of those ellipses (e.g axial ratio) as tabular features. This was chosen to create features which were sensitive to the shape of galaxies (LB21). In this work, 
we instead use the DECaLS model as a feature extractor, with the learned representation forming the features for each galaxy.
We believe our learned representation is particularly vital for tasks where the interesting morphology (e.g. irregular shapes, rings) cannot be well-described as a series of ellipses of increasing flux. 
More broadly, because the galaxies are arranged in representation space by visual similarity (Sec. \ref{sec:representation_evidence}), interesting galaxies are likely to have similar representations and so representations are a useful feature for predicting user interest.

Next, we change the regressor modelling user interest. We replace Baseline's Random Forest with a Gaussian Process (GP, \citealt{Rasmussen2006}).
Gaussian Processes define a probability distribution over possible functions. The space of possible functions is set by the choice of kernel, $\kappa(x, x')$. The kernel defines an effective distance between points, with the range of probable values for each point being constrained by the values of known nearby points. The kernel hyperparameters (e.g. the typical distance over which known points have a strong constraining effect) are fit to maximise the likelihood of the observed (training) data. See \cite{Murphy2012} for a concise review and \citealt{Rasmussen2006} for a comprehensive treatment.

Gaussian Processes are particularly appropriate here for two reasons. First, they can flexibly model smooth distributions; they make no parametric assumptions about the shape of the user interest distribution other than through the kernel itself. We use a rational quadratic kernel\footnote{We also find that performance is similar using a Matern kernel.} assuming user interest is similar for similar galaxies and varies over some typical scale, and add a white component to model intrinsic label uncertainty on user interest and to model noise in the underlying representation. Second, through marginalising over the many possible functions allowed by the kernel, GPs provide relatively reliable uncertainties. Indeed, GP uncertainties are sometimes considered the `gold standard' against which more scalable methods are measured \citep{Houlsby2014}. Knowing the uncertainty of our user interest predictions for each galaxy is critical for applying active learning.

A key part of active learning is the acquisition function i.e. which galaxies to label.
\texttt{Astronomaly} selects galaxies to label with a `joint' score based on both expected interest and rarity. For each galaxy, if galaxies with similar features have already been labelled, the regressor is considered more reliable and the joint score is weighted towards the regressor's predicted interest. If not, the joint score is instead weighted towards the galaxy's rarity. Users are then asked to label the galaxies with the highest joint score i.e. the galaxies thought most likely to be interesting anomalies. Whilst effective, this algorithmically greedy approach may be inefficient in some cases. We may not yet know which galaxies are likely candidates, and should therefore devote at least some labelling effort to explicitly helping the regressor model user preferences.

Which galaxies would best help model user preferences?
Active learning acquisition functions are generally concerned with modelling a function globally: in our case, modelling the user interest on \emph{all} galaxies. But here, we are specifically interested in finding the most interesting galaxies, i.e. modelling the function near its maxima. We are not concerned with whether a galaxy is very boring or merely somewhat boring. Modelling maxima is an optimisation problem, and we therefore use an acquisition function from the Bayesian optimisation literature.

 Specifically, we choose to use maximum expected improvement (`max EI') as our acquisition function. EI, introduced by \cite{Mockus1991} and further developed in \cite{Jones1998}, is calculated as:
 \begin{multline}
     EI(x) = (\mu(x) - f(x^+) - \epsilon)\Phi(\frac{\mu(x)-f(x^+)-\epsilon}{\sigma(x)}) \\ + \sigma(x)\phi(\frac{\mu(x)-f(x^+)-\epsilon}{\sigma(x)})
 \end{multline}
 where $\mu(x)$ and $\sigma(x)$ are the mean and variance of the GP modelling user interest, $f(x^+)$ is the current maximum recorded user interest, and $\Phi$ and $\phi$ are the CDF and PDF of a standard normal variable, respectively. Intuitively, EI measures the expected gain in maximum interest from a rating, $x$, given the current estimate, $\mu(x)$, and uncertainty, $\sigma(x)$, for $x$'s likely interest. $\epsilon$ is a hyperparameter balancing exploration and exploitation and is subtracted from the expected improvement, causing the algorithm to ignore gains smaller than $\epsilon$ (typically in well-explored regions) and instead explore more uncertain regions where the potential gains are still larger than $\epsilon$. $\epsilon$ is particularly important for this problem because we potentially aim to find diverse anomalies in many different regions each of high interest, rather than just the anomalies in the single region of highest interest. We find that a non-zero $\epsilon$ is crucial to avoiding occasional (10-20\%) failures where the acquired galaxies fall into a single local maxima. We choose $\epsilon  = 0.5$ throughout, representing an interest increase  of 0.5 on the  0-5 rating scale used by \texttt{Astronomaly}.

\subsection{Experiments}

\subsubsection{Galaxy Zoo 2 `Odd' Galaxies}
\label{sec:astronomaly_gz2}

 LB21 primarily demonstrate the performance of \texttt{Astronomaly} through identifying unusual galaxies in Galaxy Zoo 2. Specifically, they aim to identify the rare (approximately 1.5\%) subset of galaxies which more than 90\% of GZ2 volunteers described as `Odd' in the `Is there anything odd?' task. We repeat this demonstration with the method in this work and compare performance against `Baseline', the specific \texttt{Astronomaly} configuration used by LB21 and summarised in the previous Section. 

 Starting from the same Galaxy Zoo 2 images as LB21, we calculate representations using our DECaLS-trained CNN following the procedure described in Sec.~\ref{sec:feature_predictions}. As in Sec.~\ref{sec:feature_predictions}, we further reduce the dimensionality using Incremental PCA, in this case with 40 components preserving 98.1\% of the variation. We then use GP-based active learning to model user interest in this reduced representation. As in LB21, we simulate receiving user ratings through the \texttt{Astronomaly} interface using the recorded GZ2  `Odd' vote fraction\footnote{We use the vote fraction as released in the `Galaxy Challenge' Kaggle competition, following LB21's experimental protocol. Note that these vote fractions are not identical to the latest published GZ2 catalog \citep{Hart2016}, which we suggest for general scientific use.} scaled and binned to integers from 0-5, and consider anomalies as those galaxies with `Odd' vote fractions above 90\%. 
 
 We acquire (i.e. simulate rating for interest) galaxies in batches of 10, chosen to ensure that it takes no more than a few seconds to retrain the GP and identify the next galaxies to rate. This helps the user rate galaxies quickly and enjoy a responsive experience. Introducing batching did not reduce performance. The first batch of galaxies is chosen randomly\footnote{Experiments with selecting the first batch via Isolation Forest did not show a performance improvement.}. We rate for interest a total of 200 galaxies, matching LB21.
 
\begin{figure}
     \centering
     \includegraphics[width=\columnwidth]{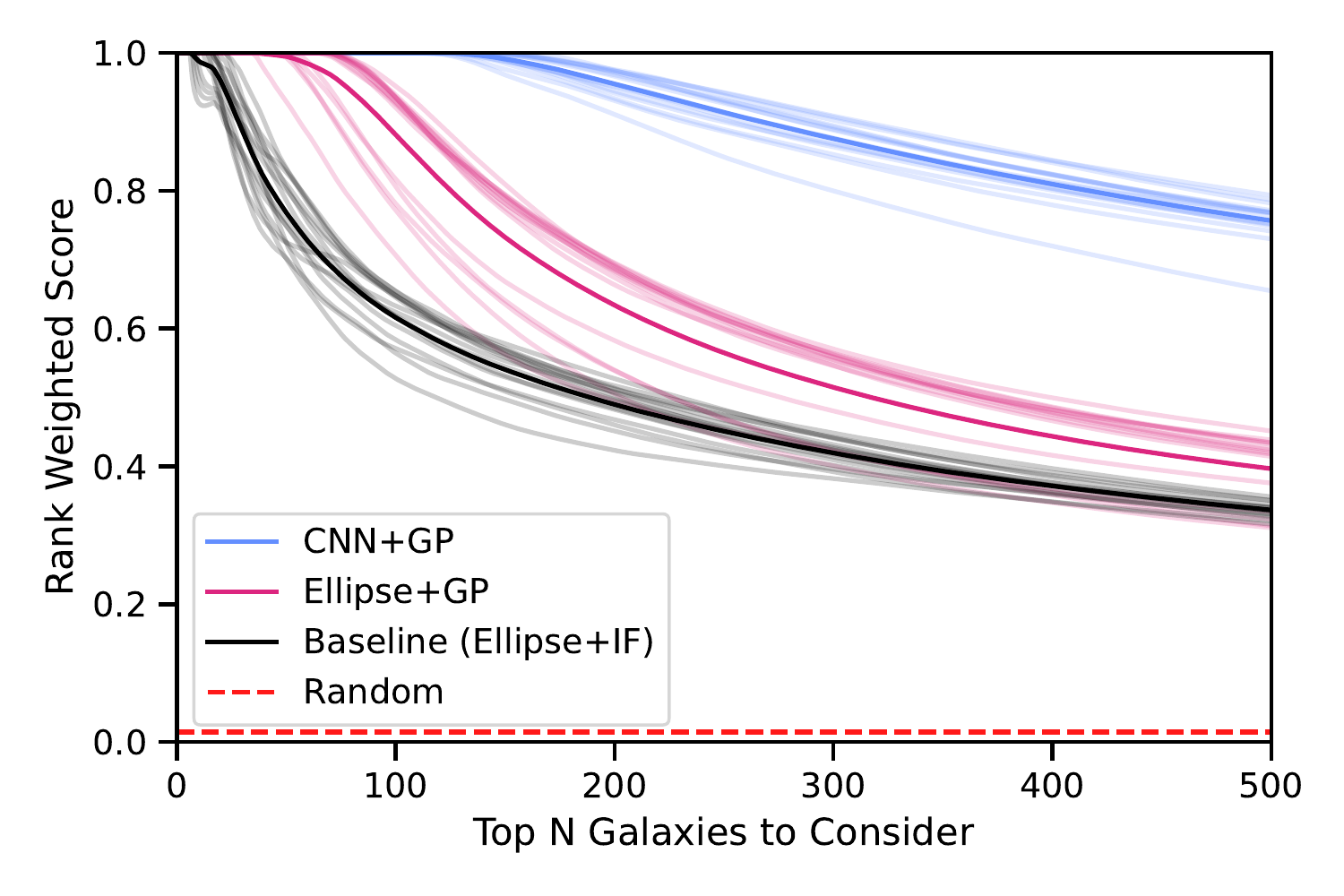}
     \includegraphics[width=\columnwidth]{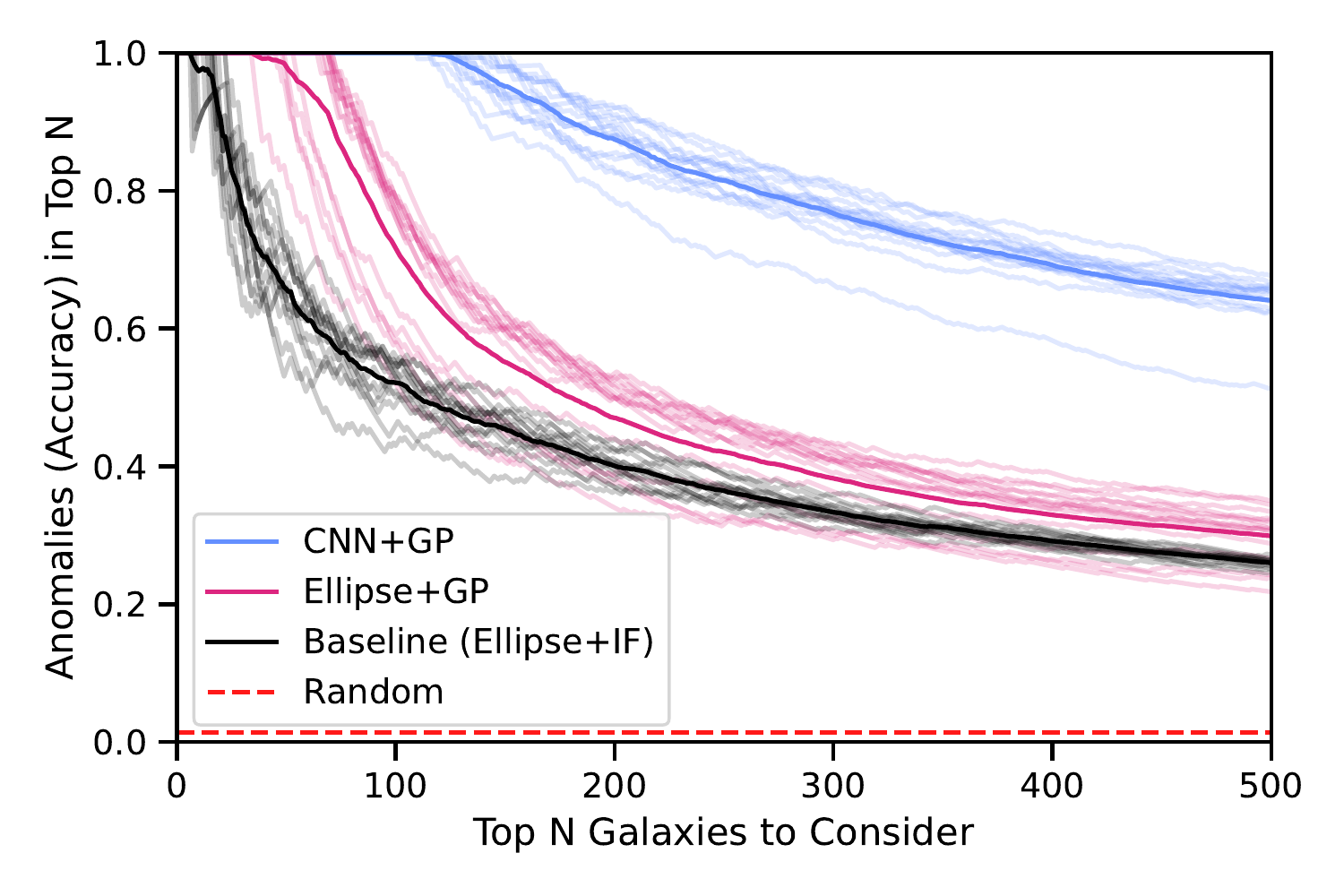}
     \caption{Rank Weighted Scores (above) and accuracy (below) for identifying `Odd' galaxies (as voted by volunteers) in GZ2 images. Calculated after training on 200 user ratings following either the method of LB21 (baseline, black) or this work (CNN \& GP, blue). The expected value from randomly-selecting galaxies is shown in red for comparison. We also show an intermediate method, using the ellipse-fitting features of Baseline and our GP active learning strategy, in magenta. Experiments are repeated 15 times, with individual runs shown as traces. The method introduced in this work dramatically improves both metrics.}
     \label{fig:gz2_results}
 \end{figure}

 Figure~\ref{fig:gz2_results} compares the results of our CNN and GP-based approach with the Baseline. Figure~\ref{fig:gz2_results} follows the same format as Figure~5 in LB21, comparing their `rank weighted score' metric against the number of top galaxies, $N$, to consider when calculating the score.
 This rank weighted score measures how highly the true interesting anomalies are ranked among the $N$ galaxies predicted as most likely to be interesting. We also provide conventional accuracy and average precision scores in Figure~\ref{fig:gz2_results} and Table~\ref{tab:metrics}. All experiments are repeated 15 times to marginalise over stochastic effects.
 
 From Figure~\ref{fig:gz2_results} it can be seen that Baseline easily outperforms random selection, with two-thirds of the 50 galaxies predicted as most likely to be interesting anomalies actually being so. Using the method presented in this work, with both CNN representations and GP-based active learning, all of the top 50 galaxies and indeed all of the top 100 galaxies are interesting anomalies. Given that interesting anomalies represent only 1.5\% of the dataset, this improvement is notable. 87\% of the top 200 galaxies are interesting anomalies, compared with 40\% using Baseline and 1.5\% using random selection.
 
\begin{figure*}
     \centering
     \hfill Ellipse Representation, Isolation Forest search \hfill      CNN Representation, GP search \hfill
     \\
     \includegraphics[width=0.45\textwidth]{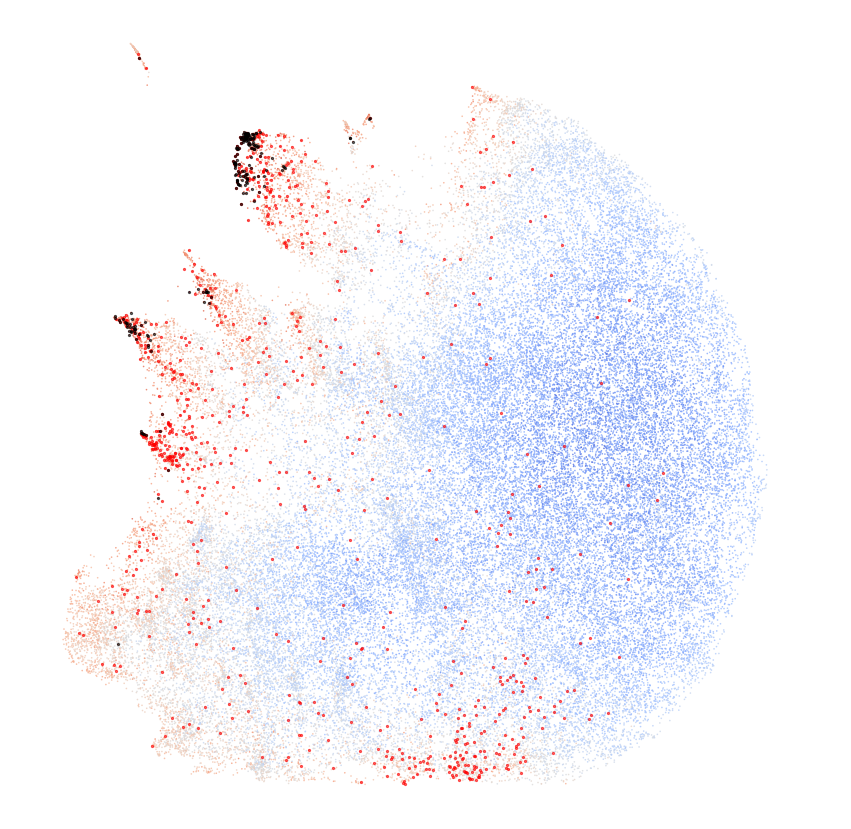}
     \hfill
     \includegraphics[width=0.45\textwidth]{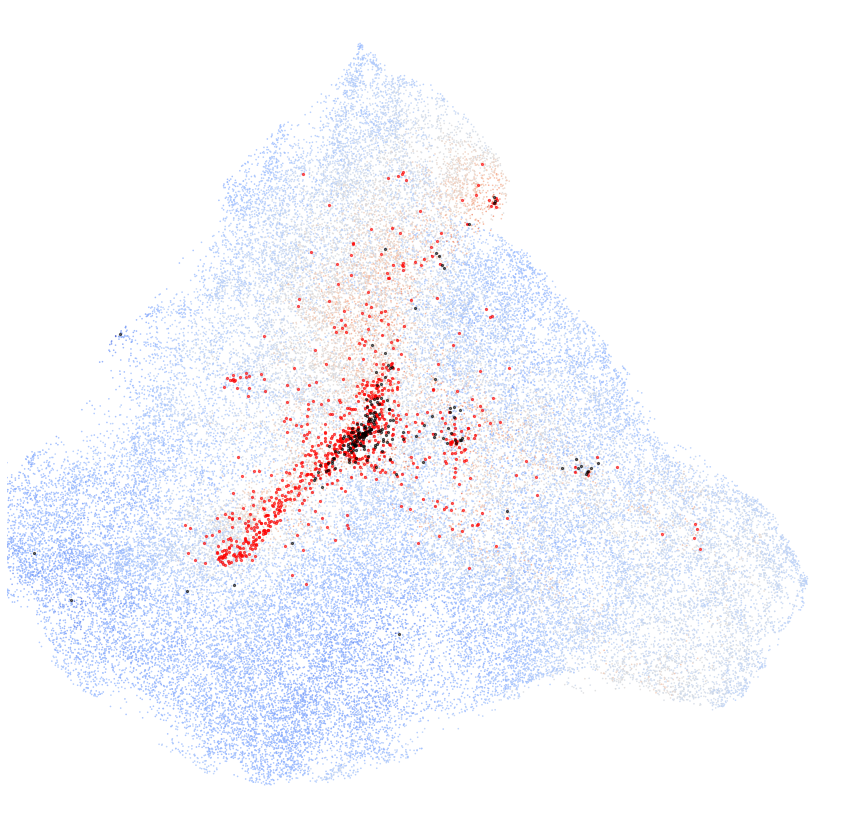}
     \caption{Visualisation of each anomaly-finding method (LB21, left, and this work, right). Small translucent points are GZ2 galaxies, coloured by the user interest predictions of each method (red for high interest, blue for low interest). The LB21 visualisation shows the initial predicted anomaly score from the Isolation Forest. Solid red points are anomalies, defined as galaxies GZ2 volunteers voted `Odd'. Solid  black points are galaxies chosen to be rated  by the user following each method, to help inform the user interest model. Our representation gathers anomalies together better (red points are more clustered in the right-hand panel) making it easier for our active learning approach to identify the part of the representation most likely to include anomalies.}
     \label{fig:gz2_representation_comparison}
 \end{figure*}

 Figure~\ref{fig:gz2_representation_comparison} investigates why the new method is more successful. In this figure, we visualise the representations of both \texttt{Astronomaly}'s ellipse-fitting method and of our CNN using a 2D \texttt{umap} projection\footnote{Note that we are using \texttt{umap} to further compress (and hence visualise) the features already extracted by each method (ellipse-fitting or our CNN) and not directly applying \texttt{umap} to the images themselves. We tested using \texttt{umap} as a feature extractor and found it under-performed \texttt{Astronomaly}'s ellipse-fitting method (though improved on random chance).}, as in Section~\ref{sec:visualizing}. We then colour galaxies according to either Isolation Forest predictions or those of the Gaussian Process interest model. We also show the galaxies considered as anomalies and the galaxies selected for rating by the user, either due to the Isolation Forest ranking or our acquisition function. The CNN representation is far more effective at grouping `Odd' galaxies together\footnote{The CNN representation may be placing `Odd' galaxies largely together because, as stated previously, most `Odd' galaxies are major mergers. An isolation forest would not work well with the CNN representation due to this effect, as the `Odd' galaxies are largely not considered quantitatively unusual.} than the ellipse representation, and this, in turn, makes user interest easier to model. The interest model matches the density of interesting anomalies well and the user-rated galaxies concentrate along the region of highest interesting anomaly density. In contrast, the ellipse representation places `Odd' galaxies along a distributed border in our visualisation. This is crucial for the success of the Isolation Forest in making an initial prioritisation (which will prefer border regions). However, the galaxies considered most anomalous by the Isolation Forest, and hence rated by the user, tend to lie only in specific patches on the border and so the user ratings of those galaxies do not efficiently measure user interest along the full anomalous border.
 
We highlight that our CNN can calculate effective feature vectors from Galaxy Zoo 2 images even though it has \textbf{never been trained on Galaxy Zoo 2 data}. The CNN was only trained on GZ DECaLS images, which are significantly deeper and of higher resolution than the Galaxy Zoo 2 images (W+22).  It is well-known that CNNs can suffer from substantial performance drops in the presence of minor domain shifts barely visible to humans (e.g. contrast adjustments, added Gaussian noise, adversarial attacks - see \citealt{Moosavi-Dezfooli2017,Hendrycks2019,Ilyas2019}), and it is therefore encouraging that the CNN representation used here remains useful across different surveys without any need for retraining. 

\subsubsection{DECaLS Mergers, Rings, and Irregular Galaxies}

 The vast majority of `Odd' GZ2 galaxies are major mergers (LB21). While scientifically valuable, major mergers may not be representative of all interesting anomalies and so mergers alone may not provide a comprehensive test of an anomaly-finding method. We therefore apply our method to finding irregular galaxies and ring galaxies in DECaLS (along with mergers again for comparison), using the vote fractions reported by GZ DECaLS volunteers.
 
 We use the same DECaLS images previously described and used in Section \ref{sec:representation_evidence}.
 We select only galaxies with at least 30 total volunteer responses\footnote{For identifying mergers, where the question was modified during GZ DECaLS (see W+22), we specifically require 30 votes for the latest version of the merger question.} to ensure reliable vote fractions.
 Of 253,286 volunteer-labelled galaxies, the \texttt{Astronomaly} ellipse method fails for 2,112 galaxies, returning \texttt{nan} features; we exclude any galaxies with failed ellipse measurements from the experiment.
 We filter to relevant galaxies using automated vote fraction prediction cuts of featured fraction > 0.6 and face-on fraction > 0.75, for a final experiment catalogue of 58,982 galaxies (56,828 for identifying mergers).
 For each class of anomaly, we choose the minimum vote fraction to be defined as an interesting anomaly such that the rate of interesting anomalies is 1.5\% (matching LB21's GZ2 experiment above); $f > 0.42$ for irregular galaxies, $f > 0.57$ for rings, and $f > 0.6$ for mergers. 
As before, we use the binned volunteer vote fraction to emulate user interest responses from 0-5.

\begin{figure}
    \centering
    \includegraphics[width=\columnwidth]{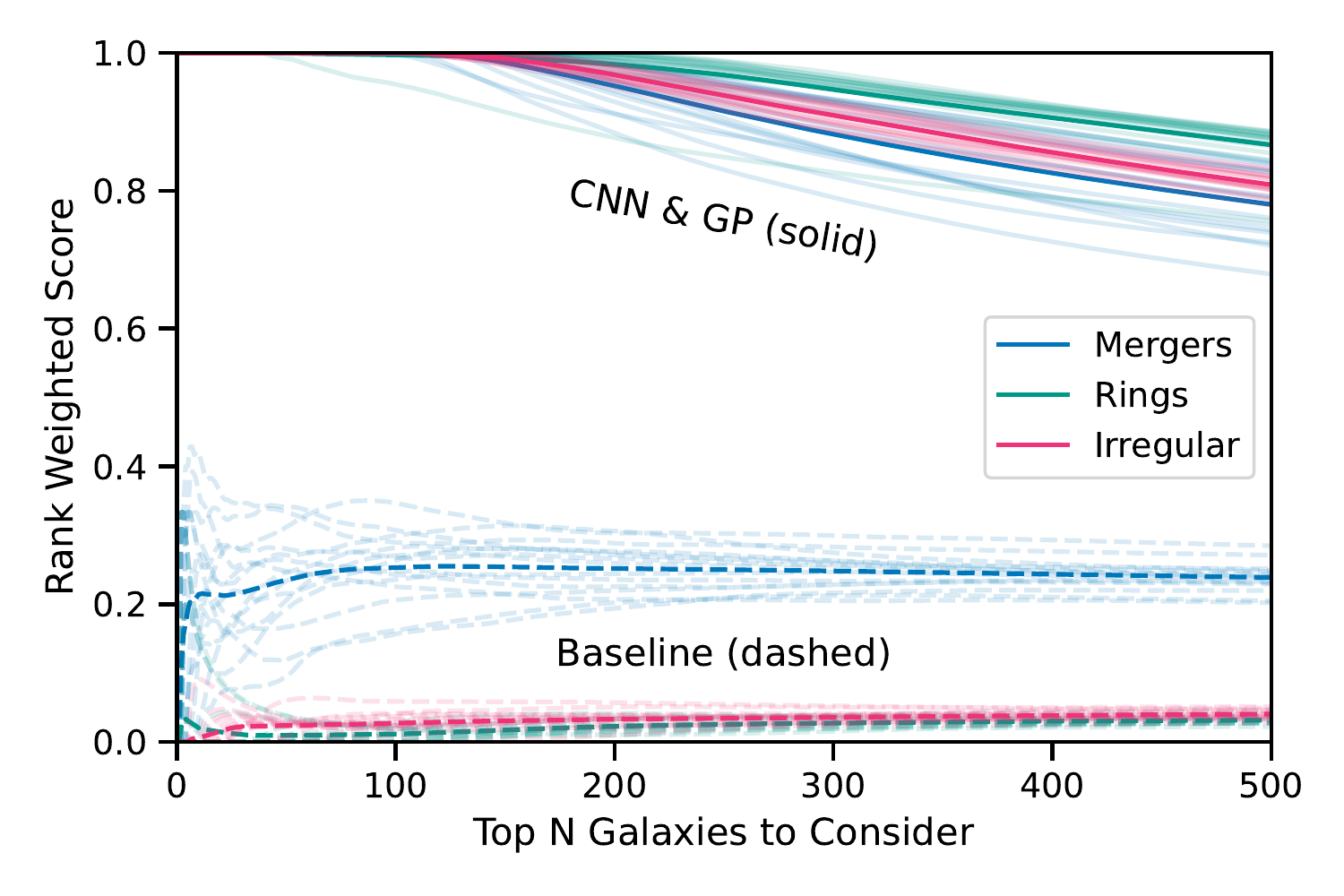}
    \includegraphics[width=\columnwidth]{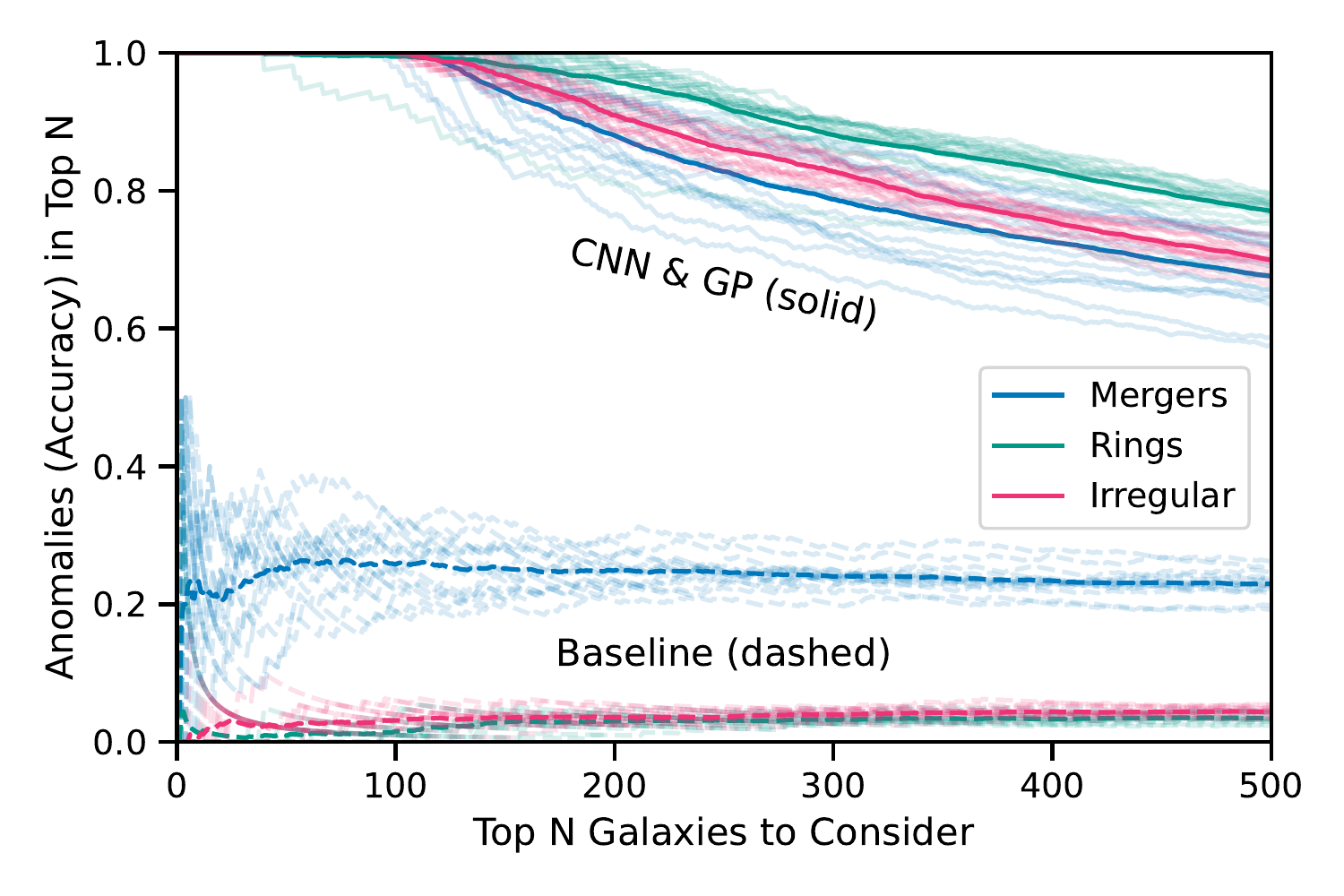}
    \caption{Rank Weighted Scores (above) and accuracy (below) for identifying interesting anomalies (mergers, rings, or irregular galaxies) in GZ DECaLS images. Calculated after training on 200 user ratings following either the method of LB21 (baseline, dashed) or this work (CNN \& GP, solid). The method introduced in this work dramatically improves both metrics.}
    \label{fig:decals_results}
\end{figure}

We follow the same method as for the GZ2 `Odd' experiment, using the CNN representation as galaxy features and acquiring galaxies (in batches of 10) that maximise the expected improvement of our GP user interest model. We compare our results to Baseline in Figure~\ref{fig:decals_results} and Table~\ref{tab:metrics}. 

Our method again dramatically outperforms both Baseline and random selection. For each anomaly class, we achieve a high fraction of true interesting anomalies in the top 200 galaxies; 88\% for mergers, 96\% for rings and 91\% for irregular galaxies. Baseline achieves 25\% for mergers and is comparable to random selection for rings and irregular galaxies. 

\begin{figure*}
    \centering
    \includegraphics[width=0.7\textwidth]{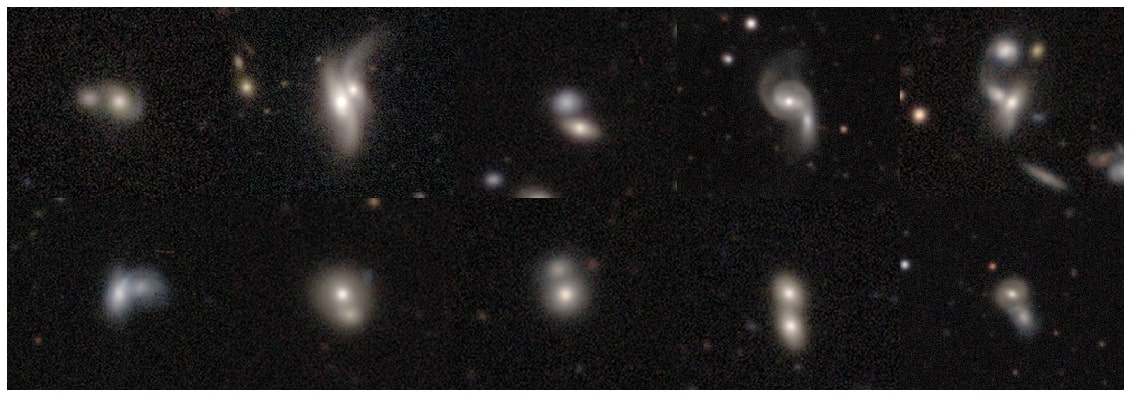}\\
    (a) Random selection from the top 200 DECaLS galaxies identified as mergers with our method
    \includegraphics[width=0.7\textwidth]{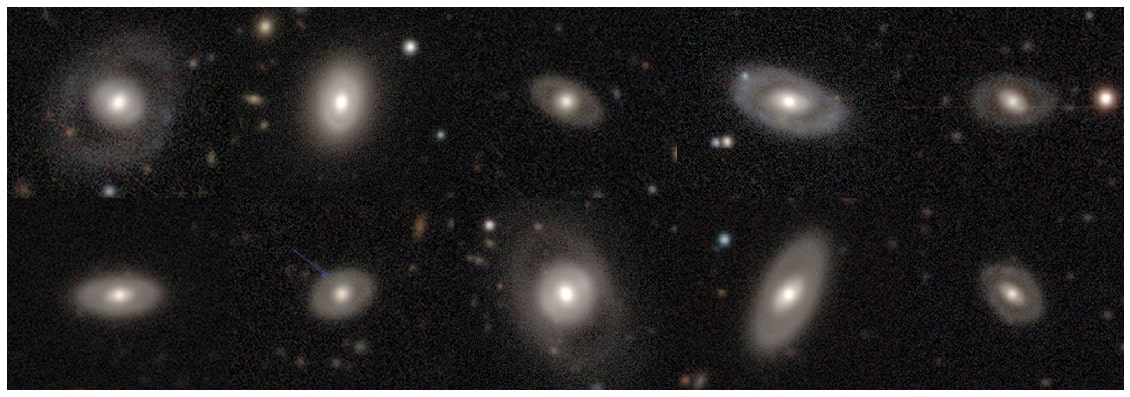}\\
    (b) Random selection from the top 200 DECaLS galaxies identified as rings with our method
    \includegraphics[width=0.7\textwidth]{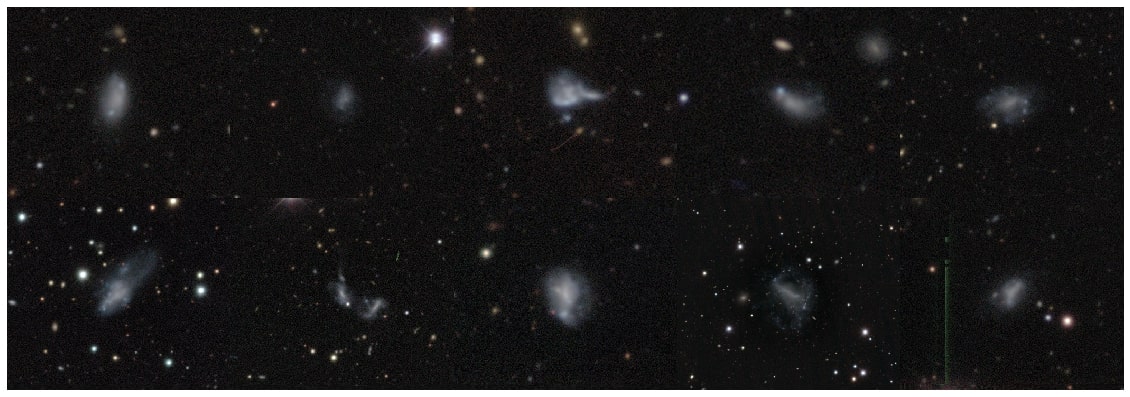}\\
    (c) Random selection from the top 200 DECaLS galaxies identified as irregulars with our method
    \caption{Random selections from the top 200 interesting anomalies identified using our CNN \& GP method, representing what a user might have found. Interesting anomalies are defined (using the GZ DECaLS vote fractions) as either mergers (upper), rings (middle), or irregular galaxies (lower).}
    \label{fig:top_decals_galaxies}
\end{figure*}

Fig.~\ref{fig:top_decals_galaxies} shows (for each anomaly class) a random selection of the top 200 galaxies identified by our method as having the highest expected interest. These are representative of the galaxies that a user being recommended interesting anomalies might see. Our method successfully presents rings, mergers or irregulars according to the user's interests. 

\begin{table*}
\centering
\begin{tabular}{l | l | l | l | l | l }
Dataset & Anomaly & Method & Average Precision & Accuracy (top 50) & Accuracy (top 200) \\
\hline
GZ2 & Odd & Baseline & 0.16 $\pm$ 0.02 & $66\% \pm 16\%$ & $40\% \pm 7\%$ \\
GZ2 & Odd & Ellipse + GP & 0.21 $\pm$ 0.10 & $99\% \pm 5\%$ & $63\% \pm 22\%$ \\
GZ2 & Odd & CNN + GP & \textbf{0.55} $\pm$ \textbf{0.10} & \textbf{100\%} & \textbf{87\%} $\pm$ \textbf{11\%} \\
\hline
DECaLS & Merger & Baseline & 0.12 $\pm$ 0.03 & 25.5\% $\pm$ 18.6\% & 24.9\% $\pm$ 7.7\% \\
DECaLS & Merger & CNN + GP & \textbf{0.58} $\pm$ \textbf{0.20} & \textbf{100\%} & \textbf{88\%} $\pm$ \textbf{15.4\%} \\
\hline
DECaLS & Ring & Baseline & Failed & Failed & Failed \\
DECaLS & Ring & CNN + GP & 0.63 $\pm$ 0.11 & 99.9\% $\pm$ 1.5\%  & 95.8\% $\pm$ 13\% \\
\hline
DECaLS & Irregular & Baseline & Failed & Failed & Failed \\
DECaLS & Irregular & CNN + GP & 0.58 $\pm$ 0.04 & 100\% & 91\% $\pm$ 5.4\% \\
\end{tabular}
\caption{Performance metrics for finding each anomaly in each dataset with either the \texttt{Astronomaly} configuration used by LB21 on this task (Baseline) or introduced in this work (CNN \& GP). `Failed' indicates a failure to find the specified anomalies, indicated by a score comparable to random selection. Errors are (roughly) estimated as the 3$\sigma$ error on the mean over 15 runs. The best metrics are shown in bold. Our method significantly improves every metric in every case, and avoids failures on some combinations of target anomaly and dataset. }
\label{tab:metrics}
\end{table*}

We have shown how our CNN representations can, when combined with GP-based modelling of user interest, be used to better find interesting optical galaxies than in previous work  -  even though it was not specifically trained to do so. In the next section, we turn to how to improve the representations themselves.

\section{Transfer Learning and Fine-Tuning}
\label{sec:transfer}

We have shown that the representations learned by our GZ DECaLS model are useful for tasks on which it was never trained.
We used the representations to find similar galaxies (Sec. \ref{sec:representation_evidence}) and interesting anomalies (Sec. \ref{sec:anomaly}) without any modification.
Going further, we can also tailor the representations for a specific task.

One relevant task is to find more of a specific type of galaxy based on a small set of known examples. We can do so better than with a similarity search by learning from multiple examples (not one), and we do not need a blind anomaly search as we know what we are looking for. By starting from our GZ DECaLS representations, we can solve this supervised classification task using far fewer known examples than otherwise required.

In this section, we fine-tune the DECaLS model for finding ring galaxies and find that it outperforms equivalent models trained from scratch, fine-tuned from ImageNet, or fine-tuned from single GZ DECaLS tasks.

\subsection{Context}
Fine-tuning is a technique where a model is trained on one problem (typically one with plentiful labelled data) and then adapted to a second problem (typically one with less labelled data).
Once trained on the first problem, the upper layers of the model (the `head') are removed and the remaining layers frozen (i.e. the weights are fixed). This `base' model simply calculates representations, exactly as we have done in Sec. \ref{sec:feature_predictions}. A new `head' model is added, with outputs appropriate to the new problem and with fewer parameters to avoid overfitting the more limited labels. The new head is trained to predict outputs for the second problem given the frozen representation (from the base model) and the new labels. This allows the new head to benefit from the previously-learned representation, as we have been doing throughout this work. Finally, once the new head is trained, some or all of the base model layers may be unfrozen and both head and base model trained together (typically at a low learning rate to avoid overfitting). This gradually adapts the representation to best solve the second problem, starting from the already-useful initial representation learned for the first problem. We refer the reader to \cite{GoodfellowBook2016} for a further introduction to fine-tuning.

What is the best base model to finetune for a new galaxy classification problem?
In our Introduction (Sec. \ref{sec:introduction}), we noted various efforts by astronomers to use fine-tuning to mitigate the lack of available labelled data for their target problem. The base models were trained either on identical narrow questions on comparable surveys \citep{Sanchez2019, Perez-Carrasco2018,Tang2019} or on the broad but terrestrial ImageNet dataset \citep{Ackermann2018, Wu2018, Martinazzo2020}. 
Training on identical questions is not possible where we want to answer new questions for which no labels yet exist.
We argued that ImageNet is qualitatively different to galaxy images and so pretraining on ImageNet is unlikely to be as helpful as pretraining to answer a broad set of questions on galaxy images. Both approaches would lead to generic representations, but ImageNet would lead to a generic \emph{terrestrial} representation while galaxy images would lead to a generic \emph{galaxy morphology} representation.
We have shown in the preceding sections that such generic galaxy representations are indeed learned and are immediately useful for diverse tasks beyond classification. We now test if our representations can help astronomers outperform ImageNet pretraining on new classification tasks.

\subsection{Experiments}
\label{sec:experiments}

To measure the effectiveness of fine-tuning from our representation to solve new classification tasks, we experiment with identifying ring galaxies.\\

Rings have long been thought to be typically\footnote{Some so-called collisional rings are caused by mergers \citep{Lavery2004}. These are thought to be rare and could be an interesting target for the similarity search in Sec. \ref{sec:similarity}.} caused by resonances in a disk driven by a bar, or, where no bar is present, driven by an oval-shape or spiral potential \citep{Schwarz1981}. More recent theoretical work suggests they may in fact be related to dynamical manifolds \citep{Athanassoula2009RingsBlocks}. Both theories predict ring morphologies broadly similar to those observed \citep{Buta2013}. However, each theory makes specific predictions about the nature and frequency of ring subtypes and so it may be possible to distinguish the true cause(s) from a sufficiently large ring sample. 

Rings are also useful to measure secular evolution. The slow nature of ring formation (in both theories) suggests a lack of recent major mergers and so any difference in characteristics between rings and standard disk galaxies may help test the effect of major mergers on topics such as quenching \citep{Smethurst2017} and black hole growth \citep{Simmons2013}. Such an investigation would again likely require a large ring sample in order to control for other variables (mass, redshift, environmental density, etc.)

Existing expert catalogues contain of order tens to hundreds of rings \citep{Buta1996, Lavery2004, Nair2010, Struck2010, Moiseev2011, Comeron2014, Buta2015,Buta2019}. This is in stark contrast to the size of modern surveys such as DECaLS \citep{Dey2018}, which contain hundreds of thousands to millions of galaxies with imaging appropriate for identifying rings. Even if rings make up only a few percent of galaxies, this suggests that there are thousands to tens of thousands of rings yet to be identified in DECaLS alone.

Efforts at automatic identification are sparse. Our literature search revealed only two papers \citep{Timmis2017, Shamir2020Rings} automatically identifying 185 and 443 ring candidates in PanSTARRS and SDSS, respectively. The largest ring catalogue, \citealt{Buta2017catalog}, was created using crowdsourcing. 3,692 galaxies were identified by Galaxy Zoo 2 volunteers and then classified by a single expert (R. Buta).

For our experiment in automatically finding rings, we first need to identify large samples of ringed and not-ringed galaxies in DECaLS images.
As mentioned in Sec. \ref{sec:anomaly}, Galaxy Zoo DECaLS volunteers were asked if each galaxy had rings via the `Are there any of these rare features?' question (see W+22 for a full schema). We use these votes to identify likely rings and not-rings.

We make initial selection cuts on the DECaLS DR5 catalogue (Sec. \ref{sec:decals_data}). For simplicity, we select galaxies with volunteer votes from the GZD-5 campaign (253,286 galaxies). We use the ML-predicted vote fractions of `Smooth' < 0.25 and `Not Edge On' > 0.75 to select a subset of 82,898 candidate ring galaxies\footnote{Our volunteer `ring' vote fraction criterion is less reliable for galaxies failing these cuts; galaxies with volunteer  `ring' vote fractions $f$ > 0.25 which are also predicted to be extremely smooth or edge-on are often judged to not be rings by the authors.}. Note that we use the ML-predicted morphology vote fractions, rather than the volunteer vote fractions, because we ultimately hope to make the same selection cuts on galaxies not previously classified by humans. We then use volunteer `ring' vote fractions to select relatively clean samples of rings and not-rings. Based on inspection by one of the authors (MW) of several hundred random galaxy images selected at various `ring' vote fractions, we choose to consider galaxies with a fraction $f > 0.25$ as rings (12\%, N=9,947; of which we judge approximately 90\% are truly ringed based on random inspection) and galaxies with a vote  fraction $f < 0.05$ as `not rings' (61\%, N=50,855). Galaxies with intermediate vote fractions (27\%, N=22,096) are discarded. Our priority is to make a simple, reliable test of how different methods perform at finding rings under equivalent conditions rather than finding as many rings as possible. We note that, with approximately 10,000 likely rings, this is the largest ring catalogue to date.

What is the best way to train a model to find these ring galaxies? We test three training methods.
First, and most conventionally, training from scratch using a random weight initialisation (`scratch').
Second, training a new head on a pretrained base model with frozen weights (`frozen').
Third, once the new head has been trained, allowing some or all of the base model layers to also be trained (`fine-tuned').
We test pretraining with either GZ DECaLS (i.e. using our representation as a starting point) or with ImageNet. This allows us to measure whether our GZ DECaLS representation is helpful, whether it is more helpful than the ImageNet representation, and whether further fine-tuning of either representation can improve performance. To investigate whether learning to solve many diverse tasks is important for creating a helpful representations, we also test pretraining with GZ DECaLS labels but only using the labels from a single task (e.g. only training to predict `Smooth or Featured?' votes).

We ensure that, other than the different training methods described above, all other factors are equivalent between tests. Below, we describe the specific details of our architecture, data splits, and training procedure.

\subsubsection{Architecture}

We use the same EfficientNetB0 architecture and training procedure as previously introduced in Sec. \ref{sec:feature_predictions}. We instantiate the network in three ways: (i) randomly, (ii) with the weights from pretraining on ImageNet as provided by Keras Applications\footnote{\href{https://keras.io/api/applications/}{https://keras.io/api/applications/}}, or (iii) with the weights from pretraining on GZ DECaLS by W+22 (released with this work, see Sec. \ref{sec:data_availability}), either pretraining to predict all GZ DECaLS tasks (as done throughout this work) or pretraining on only a single GZ DECaLS task (this section only). In all cases, we replace the final dense layer with a new head comprised of two 64-unit dense layers, each with dropout probability of $p=0.75$ and \texttt{relu} activations \citep{Agarap2018}, and a final 1-unit dense layer with sigmoid activation. This head design was chosen to have a low capacity to minimise overfitting on small datasets. For pretrained models, the head is trained to convergence on the frozen base model (using the Adam optimiser and an initial learning rate of $10^{-3}$) before the base model is unfrozen and allowed to also train (with a lower learning rate of $10^{-5}$).
Using our chosen head, EfficientNetB0 has approximately 4.1m parameters, comparable to older designs such as VGG16 \citep{Simonyan2014}, and is therefore best viewed as a more advanced network rather than a `bigger' network. 

Two elements of EfficientNet are particularly important to this work. 
First, EfficientNet includes batch normalisation layers \citep{Ioffe2015} which we never unfreeze during fine-tuning  (i.e. we preserve the activation statistics from initial training), as is standard practice. Second, EfficientNet is divided into a repeating pattern of mobile inverted bottleneck blocks, just as previous designs tend to include repeating blocks of convolutional layers and a pooling layer. We investigate partially fine-tuning EfficientNet by unfreezing increasing numbers of these blocks, from the output layer down, and measuring how performance varies. We follow the same block naming schema as \citealt{Tan2019a}'s EfficientNet implementation\footnote{\href{https://github.com/qubvel/efficientnet}{https://github.com/qubvel/efficientnet}}; the `top' block is the Conv2D and batch normalisation block listed as Stage 9 in \cite{Tan2019a} Table 1, `block7' is the mobile convolutional block listed as Stage 8, `block6' is listed as Stage 7, and so forth\footnote{The blockN and Stage N+1 numbers are offset because the implementation names the first block as `stem' rather than Stage 1.}.

\subsubsection{Restricting Dataset Size}

Not all astronomers have access to tens of thousands of labelled galaxies.
It is therefore crucial to measure how the performance of each training method varies with the number of available labels.
We expect that starting from the GZ DECaLS representation will be particularly useful for astronomers with fewer labelled galaxies, where training from scratch would be more likely to overfit.

When varying the dataset size, the class balance must remain constant regardless of dataset size so that the final losses are comparable\footnote{Class imbalance makes prediction easier. Consider the limiting case where there is only one class; a toy model predicting only that class would be perfectly accurate.}. We choose the balance to be equal. 
For each experiment run, we first set aside 30\% of rings (2,984), chosen randomly, and divide them into validation (10\%) and test (20\%) sets. We then similarly set aside 30\% of not-rings and randomly select 2,984 not-rings to match each ring. 
To construct the training set, we oversample (i.e. repeat) the remaining 6,962 rings by a factor of 5 such that the number of remaining ringed galaxies is close to, but slightly below, the number of not-ringed galaxies\footnote{This also allows us to train on more non-ringed galaxies than simply picking an equal number of non-ringed galaxies, because each of the 5 repeats of a ring galaxy are matched by a unique non-ring galaxy.}. We then cut surplus not-ringed galaxies such that the class balance is exactly equal (6,962 rings repeated five times each, and 34,810 unique not-rings). We then artificially reduce the dataset size as required for the desired dataset size by dropping random galaxies from the training subset. This provides realistic variation in the training class balances while preserving the average balance. We do not drop galaxies from the validation subset; preserving these galaxies drastically reduces the noise in our performance metrics introduced by early stopping (below).

Every model is independently trained with a new train, validation and test split. This allows us to measure the significant uncertainty in loss caused by the choice of training data, particularly in the low data regime; training on \textit{these} 10 or \textit{those} 10 galaxies can dramatically affect model performance. 

\subsubsection{Training Procedure}

Models are trained using the binary cross-entropy loss. To  efficiently use our limited GPU resources, we use early stopping (i.e. we end training for models with a non-decreasing validation loss).The number of update steps per epoch increases with dataset size and so we calculate the patience (i.e. the maximum number of epochs with no \emph{validation} loss improvement before cancelling training) on a sliding scale from 10 to 30. Specifically, after some experimentation, we choose the patience as min(max(10, int(epochs/6)), 30) and the total possible epochs\footnote{No model is trained to the maximum number of epochs; this is solely used to calculate the patience for early stopping.} as $5 \times 10^6$/train dataset size. We find this ensures that all models are trained to convergence but GPU resources are not unduly wasted past convergence.
Training time is strongly dependent on dataset size. Training on the full dataset takes approximately 6 hours on an NVIDIA A100 GPU.
As our performance metric, we record the \emph{test} loss of the weights with the lowest observed \emph{validation} loss during training (i.e. the best-performing checkpoint as measured on the validation dataset).

We experiment with the following training methods.
For initial weights pretrained on all GZ DECaLS tasks, on the `Smooth or Featured' task only, or on Imagenet, we test six fine-tuning options (the top block only, blocks 7+, 6+, 5+, 4+, and all blocks), each first training atop a frozen base model before fine-tuning.
For initial weights pretrained on the GZ DECaLS `Spiral' task only, `Bar' task only, and `Bulge' task only, we test two fine-tuning options (top block only and all blocks), each first training atop a frozen base model similarly.
We also train a model from scratch.
All combinations of training method and training dataset size are repeated 5 times for each of the 12 dataset sizes, for a total of 60 models per training method. 
We record performance metrics from a total of 2,940 models.


\subsection{Results}

\begin{figure}
    \centering
    \includegraphics[width=\columnwidth]{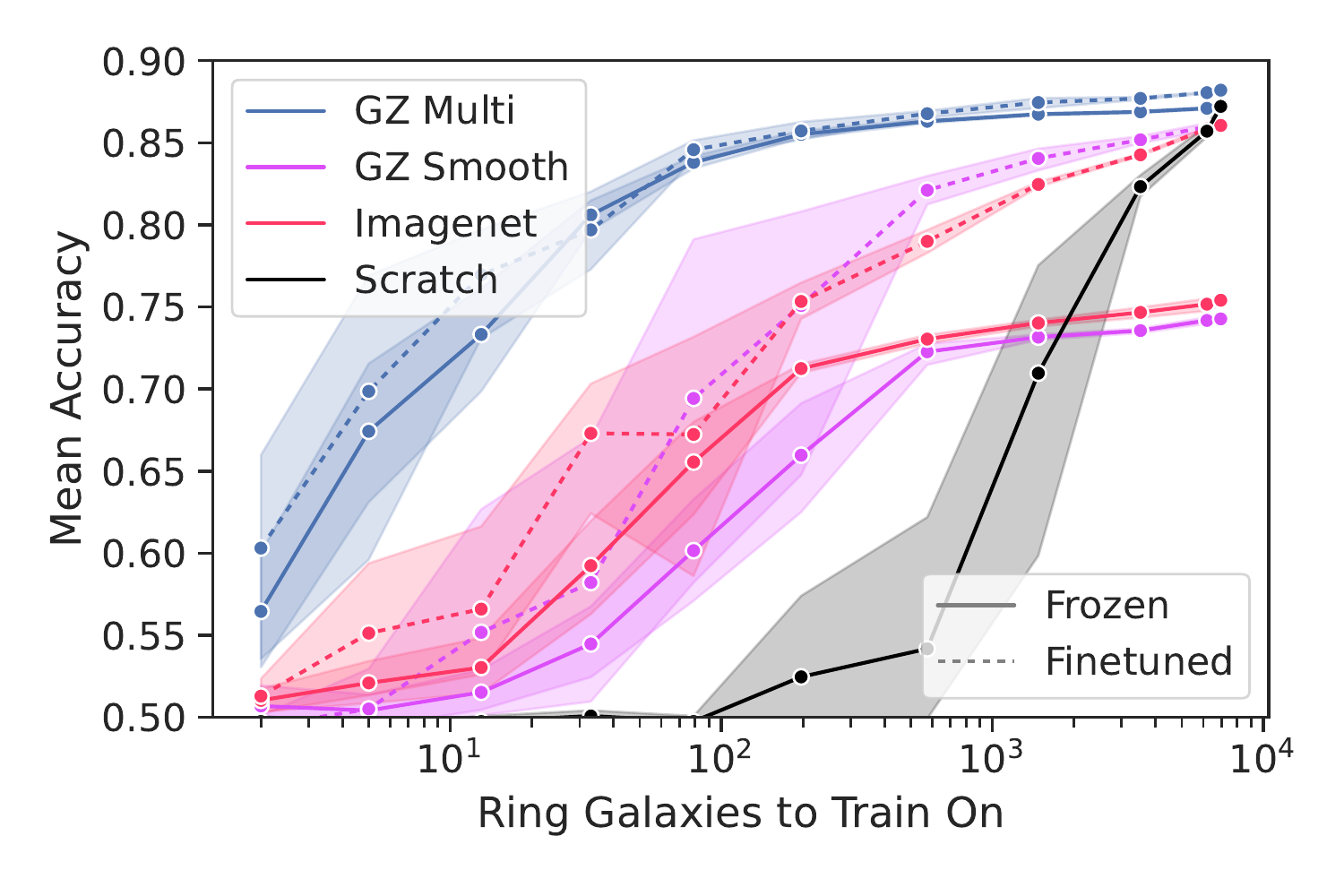}
    \caption{
    Test accuracy as a function of the total number of labelled ring galaxies to train on, split by base model pretraining.
    Models are pretrained on either all (`Multi') GZ DECaLS tasks (i.e. starting from our DECaLS representation), pretrained to solve only the Smooth/Featured/Artifact GZ DECaLS task, pretrained on ImageNet, or trained from scratch. Solid vs. dashed lines compare models where only the upper-most layers are allowed to train (`Frozen') vs. where all layers are allowed to train (`Finetuned'). Models pretrained on GZ DECaLS are better able to classify rings at all training set sizes.\\
    }
    \label{fig:accuracy_by_rings}
\end{figure}

We find that models pretrained with all GZ DECaLS tasks outperform both models pretrained with ImageNet and models trained from scratch for datasets of all available sizes. Figure \ref{fig:accuracy_by_rings} reports the mean accuracies.

For very small training datasets (below 10 unique rings), all models struggle similarly but the DECaLS-pretrained models already improve on random chance. With 10-100 rings, the DECaLS-pretrained models vastly outperform all others. With 100-1000 rings, the fine-tuned ImageNet model improves significantly but the DECaLS-pretrained models remain firmly ahead. With $10^3-10^4$ rings, training from scratch suddenly becomes feasible; the from-scratch model dramatically improves from random chance (i.e. failing to train) to outperform the ImageNet model. The DECaLS-pretrained model remains ahead with our full training dataset of 6,962 rings, though the from-scratch would likely equal or overtake it with around $10^4-10^5$ labelled rings. Since approximately 12\% of galaxies in our dataset have rings, this would correspond to labelling $10^5-10^6$ galaxies.

\begin{figure}
    \centering
    \includegraphics[width=\columnwidth]{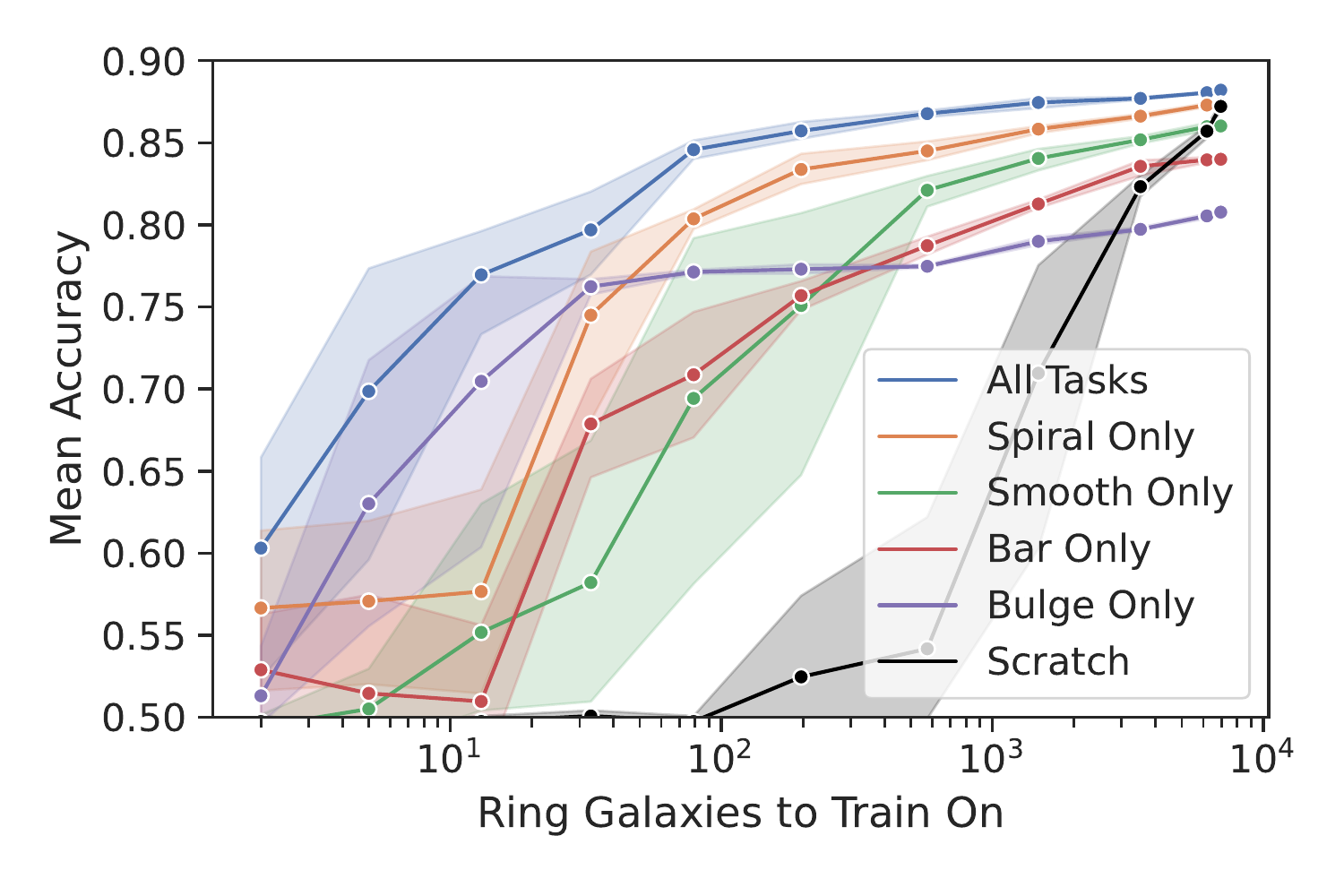}
    \caption{Test accuracy as a function of the total number of labelled ring galaxies to train on, split by base model pretraining (as Fig. 7, above), but comparing the performance of models pretrained on only one GZ DECaLS task to models pretrained to solve all GZ DECaLS tasks simultaneously. The individual GZ tasks are either Spiral Yes/No, Smooth/Featured/Artifact, Bar Strong/Weak/None, or Bulge Size. All models are finetuned. Models pretrained to solve all GZ DECaLS tasks are better able to classify rings than models pretrained on any individual GZ DECaLS task.}
    \label{fig:single_vs_all}
\end{figure}

\begin{figure*}
    \centering
    \includegraphics[width=\textwidth]{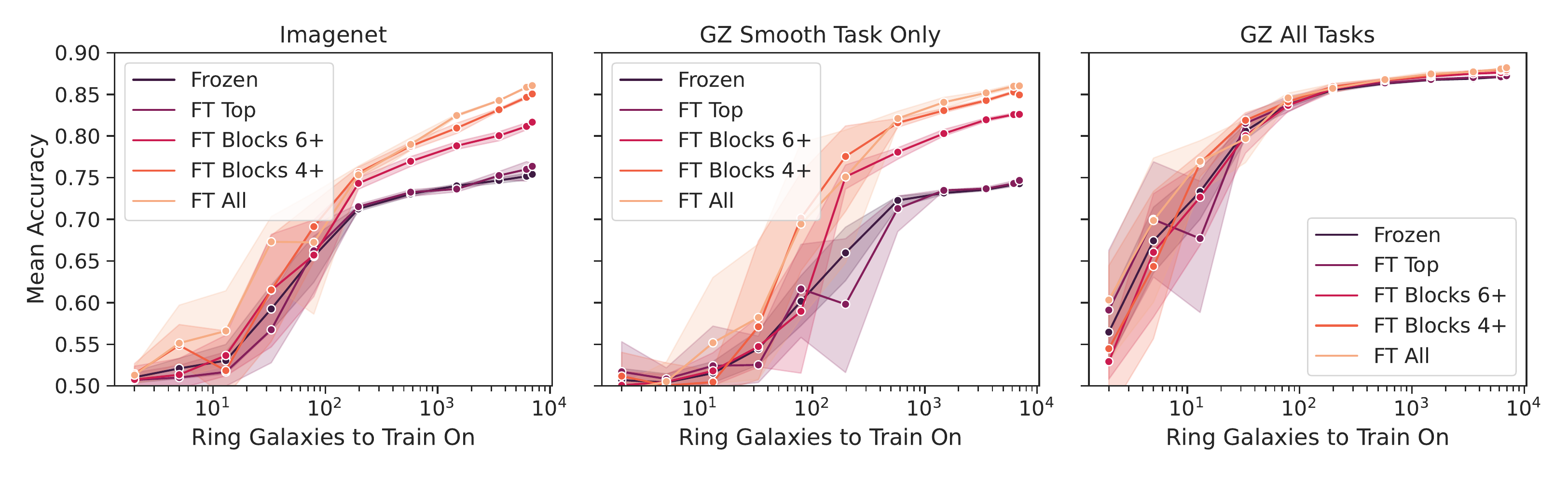}
    \caption{
        The effect of fine-tuning on ring classification test accuracy, split by base model pretraining. The model layer blocks are named, from the output, `Top',  'Block7', 'Block6', etc - see main text for details. We show results for fine-tuning the top (Top), blocks six and above (Blocks 6+), blocks four and above (Blocks 4+) and all blocks (All). For models pretrained with either ImageNet (left) or on the GZ DECaLS `Smooth or Featured' task (centre), fine-tuning intermediate layers is crucial to achieve good performance. In contrast, for models pretrained to solve all GZ DECaLS tasks simultaneously (right), performance is high without fine-tuning and fine-tuning provides only a small additional benefit. This suggests the representation learned from all DECaLS tasks is more immediately appropriate to new tasks than representations learned from single DECaLS tasks or from Imagenet.
        }
    \label{fig:finetuning_comparison}
\end{figure*}

Two further comparisons suggest that training on multiple tasks is crucial for constructing a useful representation for this new task (identifying rings). Figure \ref{fig:single_vs_all} shows that, after fine-tuning all layers to classify rings, models pretrained on all DECaLS tasks significantly outperform equivalent models pretrained on any one of several individual tasks. Figure \ref{fig:finetuning_comparison} compares the final performance of models fine-tuned to increasing depths - from only the top mobile bottleneck block (`Top'), through the intermediate blocks (Blocks 6 and above, 4 and above) and down to all layers (`All'). Fine-tuning more layers consistently improves the performance of all models, but the magnitude of this performance improvement is dramatically different. For models pretrained on either Imagenet or on the DECalS `Smooth or Featured' single task, fine-tuning only the top block has little to no effect on accuracy vs. the frozen equivalent (where only the dense layers are trained to classify rings). Fine-tuning the intermediate blocks and above is necessary to achieve good performance, increasing accuracy from approx. 70\% to approx. 85\%. In contrast, for models pretrained on all DECaLS tasks, even the frozen models outperform the fully fine-tuned Imagenet and single task models, with further fine-tuning providing only a small additional performance improvement. Together, we interpret these comparisons as strong evidence that the representations learned from training on all DECaLS tasks are more immediately appropriate to new tasks than representations learned from single DECaLS tasks or from Imagenet.

Our practical advice is that if you have fewer than $10^4$ labelled galaxies of one class, and the task you are solving is of a similar nature to the Galaxy Zoo questions, you are likely to perform better with our pretrained model than with a comparable model either trained from scratch or pretrained on ImageNet. The further gain from introducing fine-tuning (rather than simply using the frozen pretrained representation) may depend on how similar the task is to the Galaxy Zoo questions.

To help the community benefit from our pretrained models, we release the code as the Python package \texttt{zoobot} at \href{https://github.com/mwalmsley/zoobot}{https://github.com/mwalmsley/zoobot}. We provide extensive documentation at \href{https://zoobot.readthedocs.io/}{zoobot.readthedocs.io} aimed at researchers (including Masters or PhD students) with a strong interest in deep learning but no prior experience. The package additionally contains simple working examples to extend pretrained models and apply fine-tuning. We  hope this will help make deep learning accessible to astronomers working with smaller labelled datasets.

\section{Discussion}

We have shown that, using our representation, finding a range of interesting anomalous galaxies is straightforward. What should we do with this capability? Specifically, how do we build systems that lead to new scientific insights from those anomalies?  First, we would like to make human-in-the-loop anomaly recommendation available to as many interested humans as possible. Citizen scientists have repeatedly driven discoveries of unique objects or new classes \citep{Lintott2019Book}. We hope to make methods like the one presented here available to them on the Zooniverse citizen science platform. This might also enable us to exploit the shared interests of the crowd through recommendation engines. We could make predictions like `people with similar interests to you also liked \emph{this} galaxy'. We would also to like to encourage formal collaboration with observatories for follow-up, which - given the size of new surveys - may become the limiting factor.

The anomalies we find will depend strongly on our choice of representation.
Our DECaLS CNN representation is learned directly from data and so might be considered more flexible than handcrafted parametric feature extractors like ellipse fitting, which assume a particular functional form (e.g. that a galaxy can be described as a series of ellipses of increasing total flux). 
However, our CNN representation will have its own assumptions (from e.g. the choice of convolution sizes) and these are perhaps harder to identify than with parametric feature extractors.
There is likely no single `best' representation.
Similarly, there is likely no single best active learning strategy. Our  Gaussian Process search and acquisition function assume that the user has degrees of interest (for example, that a user interested in rings will find disks a little interesting, face-on-disks more interesting, and rings most interesting) so that our initially-random search can move up the resulting interest gradients to find the target galaxies. 
A user who only wants to find anomalies which are utterly distinct from other galaxies might be better served by starting from a machine-learning-prioritised list of the most quantitatively-unusual galaxies, as in e.g. LB21. Frameworks like \texttt{Astronomaly} are therefore important for enabling researchers to choose their own representations and active learning strategies while abstracting away shared technical details such as the browser interface. Our pretrained CNN is publicly available (Sec. \ref{sec:data_availability}) and we plan on incorporating it into future versions of \texttt{Astronomaly}.

Counter-intuitively, our fine-tuning results show that pretraining on ImageNet can actively harm performance. 
It has been previously assumed \citep[e.g.][]{Dobbels2019} that pretraining on ImageNet is good practice for astronomers. 
Evidence for the benefits of ImageNet pretraining was gathered with experiments on relatively small datasets \citep{Ackermann2018, Martinazzo2020}, where the benefits of pretraining would be expected to be greater.
We find that ImageNet pretraining is indeed useful with small datasets (consistent with those previous experiments) but that as the dataset size increases, ImageNet pretraining may eventually lead to worse performance than training from scratch. Recent computer science results show that the network initialization can dramatically change the minima found at convergence \citep{Fort2019}. It may be that, beyond a certain dataset size, the ImageNet initialisation sets the network on a path to learn features which are ultimately less helpful than those that would otherwise be learned directly. We encourage researchers to experiment with both our galaxy-appropriate pretrained weights and with training from scratch.

When aiming to solve a new task on galaxy images, it may seem self-evident, in retrospect, that pretraining on galaxy images is more effective than pretraining on terrestrial images. Doing so is not standard practice. We are the first to do this in a supervised context.
It remains to be seen whether pretraining on large supervised datasets or on even larger self-supervised datasets will be more effective for astronomers. Self-supervised contrastive learning in particular continues to advance \citep{Grill2020,Caron2021EmergingTransformers} and has found very recent practical applications in astronomy \citep{Hayat2021self, Sarmiento2021, Stein2021}. One key benefit in our context is the opportunity to distinguish classes where broad supervised labels are unhelpful (for example, we note that the similarity search returns two `wrong size' images under the `star' search, perhaps because the GZ decision tree labels are not useful in distinguishing these two cases). The best approach may ultimately be a combination of the two, with self-supervised `pre-pretraining' followed by supervised pretraining and finally fine-tuning to the task at hand. As the complexity grows, we may see some ML-centric researchers specialise in this process while others focus on carrying out specific applications and drawing scientific conclusions. 

Bias propagation is a potential concern. The initial representation could include subtle unwanted correlations from the initial dataset, which could then be inherited by the fine-tuned model and affect the downstream predictions even if the downstream dataset is itself unbiased. Detecting such biases is itself an active field of computer science research and so this, again, would likely benefit from specialised attention. We note that only a small minority of recent deep learning applications in the astronomical literature include experiments to check for biases or to understand why specific predictions are made (e.g. \citealt{Ghosh2020}). This might either make inherited bias more of a concern, in that such biases are unlikely to be detected without a change in culture within the field, or less of a concern, in that the field currently broadly accepts models with a potential direct bias and so inherited bias is merely a second-order effect on an existing issue. 

The question of how best to classify a new survey with no labels becomes increasingly pressing as we approach first light for the Vera Rubin Observatory \citep{LSSTScienceCollaboration2009} and Euclid \citep{Laureijs2011}.
We noted (Sec. \ref{sec:anomaly}) that our model was able to identify `Odd' galaxies in Galaxy Zoo 2 despite only being trained on DECaLS.
It is plausible (see e.g. \citealt{Sanchez2019}) that retraining on a few hundred survey-specific labels might have further improved performance. 
If, as our results suggest, CNN trained on broad tasks are relatively transferable between galaxy surveys, we may be able to use these pretrained models to rapidly classify entirely new surveys like LSST.
A short citizen science campaign answering a Galaxy-Zoo-like task on an active-learning-selected subset could provide enough new labels to adapt our  representations.
The adapted representations could then be used for time-sensitive downstream tasks like finding difference-image transients for which one cannot afford to wait several years to build the datasets that traditional `from scratch' approaches require.

If large datasets become central to deep learning in astronomy, as they have in the natural language community, it is crucial that we allow fair access for researchers at less well-resourced institutions.
The data, code and trained models from this work are all publicly available (Sec. \ref{sec:data_availability}).

\section{Conclusion}

Representations are the core of many machine learning methods. We have shown that, when trained on the broad task of answering every Galaxy Zoo DECaLS question, our convolutional neural network learns an internal representation of galaxies arranged by morphological similarity. This general representation is directly useful for new practical tasks on which the network was never trained.

The first practical task we showed was a similarity search method. Similarity searches aim to return the most similar galaxies to a query galaxy. To do so, they require a quantified measurement of the similarity between two galaxies - a problem underlying the search for automatic taxonomies of galaxies. Because our representation arranges galaxies by similarity, our measurement of similarity is simply estimated as the distance in representation space between galaxies. The most similar galaxies are the query galaxy's nearest neighbours. We tested our search method using free text hashtags (e.g. `\#starforming', `\#disturbed', etc.) from the Galaxy Zoo forum. For each common hashtag, we searched for the galaxies most similar to the galaxy most frequently given that tag. Our searches were successful in a clear majority of cases, even where the hashtag was highly specific (e.g. `\#dustlane') and even though none of the tags corresponded to the original training labels (Galaxy Zoo vote fractions).

The second practical task was finding rare galaxies that were personally interesting to a given user. We used Gaussian Process regression to model user interest, and our uncertainty about that interest, for each galaxy. We selected which galaxies to be rated for interest by our user with active learning and a Bayesian optimisation acquisition function. We simulated a user expressing their interest through the \texttt{Astronomaly} interface \citep{Lochner2021} with Galaxy Zoo vote fractions as a ground truth for interest values. We carefully replicated the test originally used to demonstrate a specific \texttt{Astronomaly} configuration and achieved a significant improvement in performance. All of the top 100 galaxies predicted by our method to be most interesting were voted `Odd' by Galaxy Zoo 2 volunteers. We then carried out comparable tests to identify mergers, rings and irregular galaxies in the DECaLS survey and found similarly improved results. Our method successfully identified each class of anomaly for our simulated user.

The third practical task was fine-tuning a convolutional neural network to solve a new galaxy classification task; specifically, to find ringed galaxies in DECaLS. We experimented with training the same architecture (EfficientNetB0) in three ways: from scratch, fine-tuned from ImageNet, and fine-tuned from Galaxy Zoo DECaLS (i.e. from our representations). In each case, we measured how the test loss varied with training dataset size. We found that fine-tuning from Galaxy Zoo DECaLS performed best, especially when few labelled rings were available (as would typically be the case for a new task). We therefore suggest that researchers use our pretrained weights rather than ImageNet weights for models aiming to solve new galaxy classification tasks.

Together, solving these tasks demonstrates the utility that an appropriate representation can have. We believe the future of machine learning for galaxy morphology lies in the thoughtful creation and sharing of representations, so that researchers can build on top of one another's models rather than creating them from scratch for each new problem.

\section{Data Availability Statement}
\label{sec:data_availability}

To help the community benefit from our pretrained models, we release much of the code from this work as the documented Python package `zoobot' at \href{https://github.com/mwalmsley/zoobot}{https://github.com/mwalmsley/zoobot}. This includes code for training the Galaxy Zoo DECaLS models from scratch, calculating the representations of new galaxies, and fine-tuning the trained model to new classification problems. The repository also includes the weights of the trained model used in this work.

We release the remaining code for this work at \href{https://github.com/mwalmsley/morphology-tools}{https://github.com/mwalmsley/morphology-tools} for reproducibility and future extension. This includes code for our similarity searches, our human-in-the-loop anomaly detection method, and our fine-tuning experiments.

The Galaxy DECaLS images and Galaxy Zoo DECaLS volunteer votes (with the exception of votes to the final multiple-choice question) were previously made available by \cite{Walmsley2022decals}. The `ring' multiple-choice answers are currently being used as part of a rigorous follow-up search for ringed galaxies in the DESI Legacy Surveys, powered by a combination of citizen science and deep learning. The complete ring catalogue will be publicly available. 

\section{Acknowledgements}

The data in this paper are the result of the efforts of the Galaxy Zoo volunteers, without whom none of this work would be possible. Their efforts are individually acknowledged at \href{http://authors.galaxyzoo.org}{http://authors.galaxyzoo.org}. 

MW and AMS gratefully acknowledge support from the Alan Turing Institute, grant reference EP/V030302/1. LF and KBM acknowledge partial support from the US National Science Foundation grant IIS-2006894.
ML and VE acknowledge support from South African Radio Astronomy
Observatory and the National Research Foundation (NRF) towards this research. Opinions expressed and
conclusions arrived at, are those of the authors and are not necessarily to be attributed to the NRF.

This publication uses data generated via the Zooniverse.org platform, development of which is funded by generous support, including a Global Impact Award from Google, and by a grant from the Alfred P. Sloan Foundation.

This work used IRIS computing resources funded by the Science and Technology Facilities Council.

This research made use of the open-source Python scientific computing ecosystem, including SciPy \citep{Jones2001}, Matplotlib \citep{Hunter2007}, scikit-learn \citep{Pedregosa2012}, scikit-image \citep{VanderWalt2014} and Pandas \citep{McKinney2010}. We also used the active learning library ModAL \citep{Danka2018} for our expected improvement acquisition function.
This research made use of Astropy, a community-developed core Python package for Astronomy \citep{TheAstropyCollaboration2018}.
This research made use of TensorFlow \citep{Abadi2015}.

We would like to thank Bruce Bassett and George Stein for useful comments on the draft.

We would like to thank Dustin Lang for creating the \href{wwww.legacysurvey.org}{legacysurvey.org} cutout service and for contributing image processing code. 

The Legacy Surveys consist of three individual and complementary projects: the Dark Energy Camera Legacy Survey (DECaLS; NSF's OIR Lab Proposal ID \# 2014B-0404; PIs: David Schlegel and Arjun Dey), the Beijing-Arizona Sky Survey (BASS; NSF's OIR Lab Proposal ID \# 2015A-0801; PIs: Zhou Xu and Xiaohui Fan), and the Mayall z-band Legacy Survey (MzLS; NSF's OIR Lab Proposal ID \# 2016A-0453; PI: Arjun Dey). DECaLS, BASS and MzLS together include data obtained, respectively, at the Blanco telescope, Cerro Tololo Inter-American Observatory, The NSF's National Optical-Infrared Astronomy Research Laboratory (NSF's OIR Lab); the Bok telescope, Steward Observatory, University of Arizona; and the Mayall telescope, Kitt Peak National Observatory, NSF's OIR Lab. The Legacy Surveys project is honored to be permitted to conduct astronomical research on Iolkam Du'ag (Kitt Peak), a mountain with particular significance to the Tohono O'odham Nation. The NSF's OIR Lab is operated by the Association of Universities for Research in Astronomy (AURA) under a cooperative agreement with the National Science Foundation.

This project used data obtained with the Dark Energy Camera (DECam), which was constructed by the Dark Energy Survey (DES) collaboration. Funding for the DES Projects has been provided by the U.S. Department of Energy, the U.S. National Science Foundation, the Ministry of Science and Education of Spain, the Science and Technology Facilities Council of the United Kingdom, the Higher Education Funding Council for England, the National Center for Supercomputing Applications at the University of Illinois at Urbana-Champaign, the Kavli Institute of Cosmological Physics at the University of Chicago, Center for Cosmology and Astro-Particle Physics at the Ohio State University, the Mitchell Institute for Fundamental Physics and Astronomy at Texas A\&M University, Financiadora de Estudos e Projetos, Fundacao Carlos Chagas Filho de Amparo, Financiadora de Estudos e Projetos, Fundacao Carlos Chagas Filho de Amparo a Pesquisa do Estado do Rio de Janeiro, Conselho Nacional de Desenvolvimento Cientifico e Tecnologico and the Ministerio da Ciencia, Tecnologia e Inovacao, the Deutsche Forschungsgemeinschaft and the Collaborating Institutions in the Dark Energy Survey. The Collaborating Institutions are Argonne National Laboratory, the University of California at Santa Cruz, the University of Cambridge, Centro de Investigaciones Energeticas, Medioambientales y Tecnologicas-Madrid, the University of Chicago, University College London, the DES-Brazil Consortium, the University of Edinburgh, the Eidgenossische Technische Hochschule (ETH) Zurich, Fermi National Accelerator Laboratory, the University of Illinois at Urbana-Champaign, the Institut de Ciencies de l'Espai (IEEC/CSIC), the Institut de Fisica d'Altes Energies, Lawrence Berkeley National Laboratory, the Ludwig-Maximilians Universitat Munchen and the associated Excellence Cluster Universe, the University of Michigan, the National Optical Astronomy Observatory, the University of Nottingham, the Ohio State University, the University of Pennsylvania, the University of Portsmouth, SLAC National Accelerator Laboratory, Stanford University, the University of Sussex, and Texas A\&M University.

The Legacy Survey team makes use of data products from the Near-Earth Object Wide-field Infrared Survey Explorer (NEOWISE), which is a project of the Jet Propulsion Laboratory/California Institute of Technology. NEOWISE is funded by the National Aeronautics and Space Administration.

The Legacy Surveys imaging of the DESI footprint is supported by the Director, Office of Science, Office of High Energy Physics of the U.S. Department of Energy under Contract No. DE-AC02-05CH1123, by the National Energy Research Scientific Computing Center, a DOE Office of Science User Facility under the same contract; and by the U.S. National Science Foundation, Division of Astronomical Sciences under Contract No. AST-0950945 to NOAO.




\bibliographystyle{mnras}
\bibliography{bibliography}

\begin{thebibliography}{}
\makeatletter
\relax
\def\mn@urlcharsother{\let\do\@makeother \do\$\do\&\do\#\do\^\do\_\do\%\do\~}
\def\mn@doi{\begingroup\mn@urlcharsother \@ifnextchar [ {\mn@doi@}
  {\mn@doi@[]}}
\def\mn@doi@[#1]#2{\def\@tempa{#1}\ifx\@tempa\@empty \href
  {http://dx.doi.org/#2} {doi:#2}\else \href {http://dx.doi.org/#2} {#1}\fi
  \endgroup}
\def\mn@eprint#1#2{\mn@eprint@#1:#2::\@nil}
\def\mn@eprint@arXiv#1{\href {http://arxiv.org/abs/#1} {{\tt arXiv:#1}}}
\def\mn@eprint@dblp#1{\href {http://dblp.uni-trier.de/rec/bibtex/#1.xml}
  {dblp:#1}}
\def\mn@eprint@#1:#2:#3:#4\@nil{\def\@tempa {#1}\def\@tempb {#2}\def\@tempc
  {#3}\ifx \@tempc \@empty \let \@tempc \@tempb \let \@tempb \@tempa \fi \ifx
  \@tempb \@empty \def\@tempb {arXiv}\fi \@ifundefined
  {mn@eprint@\@tempb}{\@tempb:\@tempc}{\expandafter \expandafter \csname
  mn@eprint@\@tempb\endcsname \expandafter{\@tempc}}}

\bibitem[\protect\citeauthoryear{Abadi et~al.,}{Abadi et~al.}{2016}]{Abadi2015}
Abadi M.,  et~al., 2016, {TensorFlow: Large-Scale Machine Learning on
  Heterogeneous Distributed Systems}, \url {http://arxiv.org/abs/1603.04467}

\bibitem[\protect\citeauthoryear{Abd El~Aziz, Selim  \& Xiong}{Abd El~Aziz
  et~al.}{2017}]{AbdElAziz2017}
Abd El~Aziz M.,  Selim I.~M.,   Xiong S.,  2017, \mn@doi [Scientific Reports]
  {10.1038/s41598-017-04605-9}, 7, 1

\bibitem[\protect\citeauthoryear{Ackermann, Schawinski, Zhang, Weigel  \&
  Turp}{Ackermann et~al.}{2018}]{Ackermann2018}
Ackermann S.,  Schawinski K.,  Zhang C.,  Weigel A.~K.,   Turp M.~D.,  2018,
  \mn@doi [Monthly Notices of the Royal Astronomical Society]
  {10.1093/mnras/sty1398}, 479, 415

\bibitem[\protect\citeauthoryear{Agarap}{Agarap}{2018}]{Agarap2018}
Agarap A.~F.,  2018, {Deep Learning using Rectified Linear Units (ReLU)}, \url
  {http://arxiv.org/abs/1803.08375}

\bibitem[\protect\citeauthoryear{Aggarwal, Hinneburg  \& Keim}{Aggarwal
  et~al.}{2001}]{Aggarwal2001}
Aggarwal C.~C.,  Hinneburg A.,   Keim D.~A.,  2001, in International Conference
  on Database Theory. pp 420--434

\bibitem[\protect\citeauthoryear{Albareti et~al.,}{Albareti
  et~al.}{2017}]{Albareti2017}
Albareti F.~D.,  et~al., 2017, \mn@doi [The Astrophysical Journal Supplement
  Series] {10.3847/1538-4365/aa8992}, 233, 25

\bibitem[\protect\citeauthoryear{Ardizzone, Di~Ges{\`{u}}  \&
  Maccarone}{Ardizzone et~al.}{1996}]{Ardizzone1996}
Ardizzone E.,  Di~Ges{\`{u}} V.,   Maccarone M.~C.,  1996, \mn@doi [Vistas in
  Astronomy] {10.1016/S0083-6656(96)00023-2}, 40, 401

\bibitem[\protect\citeauthoryear{Athanassoula, Romero-G{\'{o}}mez  \&
  Masdemont}{Athanassoula et~al.}{2009}]{Athanassoula2009RingsBlocks}
Athanassoula E.,  Romero-G{\'{o}}mez M.,   Masdemont J.~J.,  2009, \mn@doi
  [Monthly Notices of the Royal Astronomical Society]
  {10.1111/j.1365-2966.2008.14273.x}, 394, 67

\bibitem[\protect\citeauthoryear{Austin et~al.,}{Austin
  et~al.}{2021}]{Austin2021}
Austin J.,  et~al., 2021, {Program Synthesis with Large Language Models}, \url
  {http://arxiv.org/abs/2108.07732}

\bibitem[\protect\citeauthoryear{Barchi et~al.,}{Barchi
  et~al.}{2020}]{Barchi2020}
Barchi P.~H.,  et~al., 2020, \mn@doi [Astronomy and Computing]
  {10.1016/J.ASCOM.2019.100334}, 30, 100334

\bibitem[\protect\citeauthoryear{Baron \& Poznanski}{Baron \&
  Poznanski}{2017}]{Baron2017}
Baron D.,  Poznanski D.,  2017, \mn@doi [Monthly Notices of the Royal
  Astronomical Society] {10.1093/MNRAS/STW3021}, 465, 4530

\bibitem[\protect\citeauthoryear{Boyajian et~al.,}{Boyajian
  et~al.}{2016}]{Boyajian2016}
Boyajian T.~S.,  et~al., 2016, \mn@doi [Monthly Notices of the Royal
  Astronomical Society] {10.1093/mnras/stw218}, 457, 3988

\bibitem[\protect\citeauthoryear{Breiman}{Breiman}{2001}]{Breiman2001}
Breiman L.,  2001, \mn@doi [Machine Learning] {10.1023/A:1010933404324}, 45, 5

\bibitem[\protect\citeauthoryear{Brown et~al.,}{Brown et~al.}{2020}]{Brown2020}
Brown T.~B.,  et~al., 2020, {Language models are few-shot learners}, \url
  {http://arxiv.org/abs/2005.14165}

\bibitem[\protect\citeauthoryear{Buncher, Sharma  \& Kind}{Buncher
  et~al.}{2020}]{Buncher2020}
Buncher B.,  Sharma A.~N.,   Kind M.~C.,  2020, {Survey2Survey: A deep learning
  generative model approach for cross-survey image mapping}

\bibitem[\protect\citeauthoryear{Buta}{Buta}{2013}]{Buta2013}
Buta R.~J.,  2013, in , Planets, Stars and Stellar Systems: Volume 6:
  Extragalactic Astronomy and Cosmology.
Springer Netherlands, pp 1--89, \mn@doi{10.1007/978-94-007-5609-0{\_}1}, \url
  {https://link.springer.com/referenceworkentry/10.1007/978-94-007-5609-0_1}

\bibitem[\protect\citeauthoryear{Buta}{Buta}{2017}]{Buta2017catalog}
Buta R.~J.,  2017, \mn@doi [Monthly Notices of the Royal Astronomical Society]
  {10.1093/MNRAS/STX1829}, 471, 4027

\bibitem[\protect\citeauthoryear{Buta \& Combes}{Buta \&
  Combes}{1996}]{Buta1996}
Buta R.,  Combes F.,  1996, Fundamental Cosmic Physics, 17, 95

\bibitem[\protect\citeauthoryear{Buta et~al.,}{Buta et~al.}{2015}]{Buta2015}
Buta R.~J.,  et~al., 2015, \mn@doi [Astrophysical Journal, Supplement Series]
  {10.1088/0067-0049/217/2/32}, 217, 32

\bibitem[\protect\citeauthoryear{Buta et~al.,}{Buta et~al.}{2019}]{Buta2019}
Buta R.~J.,  et~al., 2019, \mn@doi [Monthly Notices of the Royal Astronomical
  Society] {10.1093/mnras/stz1780}, 488, 2175

\bibitem[\protect\citeauthoryear{Cardamone et~al.,}{Cardamone
  et~al.}{2009}]{Cardamone2009}
Cardamone C.,  et~al., 2009, \mn@doi [Monthly Notices of the Royal Astronomical
  Society] {10.1111/j.1365-2966.2009.15383.x}, 399, 1191

\bibitem[\protect\citeauthoryear{Caron, Touvron, Misra, J{\'{e}}gou, Mairal,
  Bojanowski  \& Joulin}{Caron et~al.}{2021}]{Caron2021EmergingTransformers}
Caron M.,  Touvron H.,  Misra I.,  J{\'{e}}gou H.,  Mairal J.,  Bojanowski P.,
   Joulin A.,  2021, in International Conference on Computer Vision. The
  Computer Vision Foundation, Montreal, \url {http://arxiv.org/abs/2104.14294}

\bibitem[\protect\citeauthoryear{Chen, Kornblith, Norouzi  \& Hinton}{Chen
  et~al.}{2020}]{Chen2020}
Chen T.,  Kornblith S.,  Norouzi M.,   Hinton G.,  2020, 37th International
  Conference on Machine Learning, ICML 2020, PartF168147-3, 1575

\bibitem[\protect\citeauthoryear{Cheng, Huertas-Company, Conselice,
  Arag{\'{o}}n-Salamanca, Robertson  \& Ramachandra}{Cheng
  et~al.}{2021}]{Cheng2020a}
Cheng T.-Y.,  Huertas-Company M.,  Conselice C.~J.,  Arag{\'{o}}n-Salamanca A.,
   Robertson B.~E.,   Ramachandra N.,  2021, Monthly Notices of the Royal
  Astronomical Society, 503, 4446

\bibitem[\protect\citeauthoryear{Clarke, Scaife, Greenhalgh  \& Griguta}{Clarke
  et~al.}{2020}]{Clarke2020}
Clarke A.~O.,  Scaife A. M.~M.,  Greenhalgh R.,   Griguta V.,  2020, \mn@doi
  [Astronomy {\&} Astrophysics] {10.1051/0004-6361/201936770}

\bibitem[\protect\citeauthoryear{Comer{\'{o}}n et~al.,}{Comer{\'{o}}n
  et~al.}{2014}]{Comeron2014}
Comer{\'{o}}n S.,  et~al., 2014, \mn@doi [Astronomy and Astrophysics]
  {10.1051/0004-6361/201321633}, 562, 16

\bibitem[\protect\citeauthoryear{Csillaghy, Hinterberger  \& Benz}{Csillaghy
  et~al.}{2000}]{Csillaghy2000}
Csillaghy A.,  Hinterberger H.,   Benz A.~O.,  2000, \mn@doi [Information
  Retrieval] {10.1023/A:1026568809834}, 3, 229

\bibitem[\protect\citeauthoryear{Danka \& Horvath}{Danka \&
  Horvath}{2018}]{Danka2018}
Danka T.,  Horvath P.,  2018, {modAL: A modular active learning framework for
  Python}, \url {https://github.com/cosmic-cortex/modAL}

\bibitem[\protect\citeauthoryear{Das, Wong, Dietterich, Fern  \& Emmott}{Das
  et~al.}{2016}]{Das2016}
Das S.,  Wong W.-K.,  Dietterich T.,  Fern A.,   Emmott A.,  2016, in IEEE 16th
  Conference on Data Mining. Institute of Electrical and Electronics Engineers
  (IEEE), pp 853--858, \mn@doi{10.1109/icdm.2016.0102}

\bibitem[\protect\citeauthoryear{Das, Wong, Fern, Dietterich  \& Siddiqui}{Das
  et~al.}{2017}]{Das2017}
Das S.,  Wong W.-K.,  Fern A.,  Dietterich T.~G.,   Siddiqui M.~A.,  2017, in
  KDD 2017 Workshop on Interactive Data Exploration and Analytics (IDEA'17).
  \url {http://arxiv.org/abs/1708.09441}

\bibitem[\protect\citeauthoryear{Dey et~al.,}{Dey et~al.}{2019}]{Dey2018}
Dey A.,  et~al., 2019, \mn@doi [The Astronomical Journal]
  {10.3847/1538-3881/ab089d}, 157, 168

\bibitem[\protect\citeauthoryear{Dieleman, Willett  \& Dambre}{Dieleman
  et~al.}{2015}]{Dieleman2015}
Dieleman S.,  Willett K.~W.,   Dambre J.,  2015, \mn@doi [Monthly Notices of
  the Royal Astronomical Society] {10.1093/mnras/stv632}, 450, 1441

\bibitem[\protect\citeauthoryear{Dobbels, Krier, Pirson, Viaene, De~Geyter,
  Salim  \& Baes}{Dobbels et~al.}{2019}]{Dobbels2019}
Dobbels W.,  Krier S.,  Pirson S.,  Viaene S.,  De~Geyter G.,  Salim S.,   Baes
  M.,  2019, \mn@doi [Astronomy and Astrophysics]
  {10.1051/0004-6361/201834575}, 624, A102

\bibitem[\protect\citeauthoryear{Dom{\'{i}}nguez~S{\'{a}}nchez,
  Huertas-Company, Bernardi, Tuccillo  \&
  Fischer}{Dom{\'{i}}nguez~S{\'{a}}nchez et~al.}{2018}]{Sanchez2018}
Dom{\'{i}}nguez~S{\'{a}}nchez H.,  Huertas-Company M.,  Bernardi M.,  Tuccillo
  D.,   Fischer J.~L.,  2018, \mn@doi [Monthly Notices of the Royal
  Astronomical Society] {10.1093/MNRAS/STY338}, 476, 3661

\bibitem[\protect\citeauthoryear{Dominguez~Sanchez et~al.,}{Dominguez~Sanchez
  et~al.}{2019}]{Sanchez2019}
Dominguez~Sanchez H.,  et~al., 2019, \mn@doi [Monthly Notices of the Royal
  Astronomical Society] {10.1093/mnras/sty3497}, 484, 93

\bibitem[\protect\citeauthoryear{Fedus, Zoph  \& Shazeer}{Fedus
  et~al.}{2021}]{Fedus2021}
Fedus W.,  Zoph B.,   Shazeer N.,  2021, {Switch transformers: scaling to
  trillion parameter models with simple and efficient sparsity}

\bibitem[\protect\citeauthoryear{Fei-Fei, Fergus  \& Perona}{Fei-Fei
  et~al.}{2006}]{Feifei2006}
Fei-Fei L.,  Fergus R.,   Perona P.,  2006, \mn@doi [IEEE Transactions on
  Pattern Analysis and Machine Intelligence] {10.1109/TPAMI.2006.79}, 28, 594

\bibitem[\protect\citeauthoryear{Fischer, Dom{\'{i}}nguez~S{\'{a}}nchez  \&
  Bernardi}{Fischer et~al.}{2019}]{Fischer2018}
Fischer J.~L.,  Dom{\'{i}}nguez~S{\'{a}}nchez H.,   Bernardi M.,  2019, \mn@doi
  [Monthly Notices of the Royal Astronomical Society] {10.1093/mnras/sty3135},
  483, 2057

\bibitem[\protect\citeauthoryear{Fort, Hu  \& Lakshminarayanan}{Fort
  et~al.}{2019}]{Fort2019}
Fort S.,  Hu H.,   Lakshminarayanan B.,  2019, {Deep ensembles: A loss
  landscape perspective}, \url {http://arxiv.org/abs/1912.02757}

\bibitem[\protect\citeauthoryear{Ghosh, Urry, Wang, Schawinski, Turp  \&
  Powell}{Ghosh et~al.}{2020}]{Ghosh2020}
Ghosh A.,  Urry C.~M.,  Wang Z.,  Schawinski K.,  Turp D.,   Powell M.~C.,
  2020, \mn@doi [The Astrophysical Journal] {10.3847/1538-4357/ab8a47}, 895,
  112

\bibitem[\protect\citeauthoryear{Goodfellow, Bengio  \& Courville}{Goodfellow
  et~al.}{2016}]{GoodfellowBook2016}
Goodfellow I.,  Bengio Y.,   Courville A.,  2016, {Deep Learning}.
MIT Press

\bibitem[\protect\citeauthoryear{Grill et~al.,}{Grill et~al.}{2020}]{Grill2020}
Grill J.~B.,  et~al., 2020, Advances in Neural Information Processing Systems,
  2020-December

\bibitem[\protect\citeauthoryear{Hart et~al.,}{Hart et~al.}{2016}]{Hart2016}
Hart R.~E.,  et~al., 2016, \mn@doi [Monthly Notices of the Royal Astronomical
  Society] {10.1093/mnras/stw1588}, 461, 3663

\bibitem[\protect\citeauthoryear{Hayat, Stein, Harrington, Luki{\'{c}}  \&
  Mustafa}{Hayat et~al.}{2021}]{Hayat2021self}
Hayat M.~A.,  Stein G.,  Harrington P.,  Luki{\'{c}} Z.,   Mustafa M.,  2021,
  \mn@doi [The Astrophysical Journal Letters] {10.3847/2041-8213/abf2c7}, 911,
  L33

\bibitem[\protect\citeauthoryear{He, Girshick  \& Doll{\'{a}}r}{He
  et~al.}{2019}]{HeRethinking2019}
He K.,  Girshick R.,   Doll{\'{a}}r P.,  2019, in International Conference on
  Computer Vision. Seoul, Korea, pp 4918--4927

\bibitem[\protect\citeauthoryear{Hendrycks \& Dietterich}{Hendrycks \&
  Dietterich}{2019}]{Hendrycks2019}
Hendrycks D.,  Dietterich T.,  2019, in 7th International Conference on
  Learning Representations, ICLR 2019. \url
  {https://github.com/hendrycks/robustness.}

\bibitem[\protect\citeauthoryear{Henrion, Mortlock, Hand  \& Gandy}{Henrion
  et~al.}{2013}]{Henrion2013}
Henrion M.,  Mortlock D.~J.,  Hand D.~J.,   Gandy A.,  2013, \mn@doi [Springer
  Series in Astrostatistics] {10.1007/978-1-4614-3508-2{\_}8}, 1, 149

\bibitem[\protect\citeauthoryear{Hocking, Geach, Sun  \& Davey}{Hocking
  et~al.}{2017}]{Hocking2017AnLearning}
Hocking A.,  Geach J.~E.,  Sun Y.,   Davey N.,  2017, Monthly Notices of the
  Royal Astronomical Society, 473, 1108

\bibitem[\protect\citeauthoryear{Houlsby}{Houlsby}{2014}]{Houlsby2014}
Houlsby N.,  2014, PhD thesis, University of Cambridge,
  \mn@doi{10.1007/BF03167379}, \url
  {http://ezproxy.nottingham.ac.uk/login?url=http://search.proquest.com/docview/1779546086?accountid=8018%5Cnhttp://sfx.nottingham.ac.uk/sfx_local/?url_ver=Z39.88-2004&rft_val_fmt=info:ofi/fmt:kev:mtx:dissertation&genre=dissertations+&+theses&sid=ProQ:ProQue}

\bibitem[\protect\citeauthoryear{Hunter}{Hunter}{2007}]{Hunter2007}
Hunter J.~D.,  2007, \mn@doi [Computing in Science and Engineering]
  {10.1109/MCSE.2007.55}, 9, 99

\bibitem[\protect\citeauthoryear{Ilyas, Santurkar, Tsipras, Engstrom, Tran  \&
  Madry}{Ilyas et~al.}{2019}]{Ilyas2019}
Ilyas A.,  Santurkar S.,  Tsipras D.,  Engstrom L.,  Tran B.,   Madry A.,
  2019, {Adversarial examples are not bugs, they are features}, \url
  {http://arxiv.org/abs/1905.02175}

\bibitem[\protect\citeauthoryear{Ioffe \& Szegedy}{Ioffe \&
  Szegedy}{2015}]{Ioffe2015}
Ioffe S.,  Szegedy C.,  2015, in Proceedings of the 32nd International
  Conference on Machine Learning. pp 448--456, \url
  {http://arxiv.org/abs/1502.03167}

\bibitem[\protect\citeauthoryear{Jones, Schonlau  \& Welch}{Jones
  et~al.}{1998}]{Jones1998}
Jones D.~R.,  Schonlau M.,   Welch W.~J.,  1998, \mn@doi [Journal of Global
  Optimization] {10.1023/A:1008306431147}, 13, 455

\bibitem[\protect\citeauthoryear{Jones, Oliphant, Pearu  \& {Others}}{Jones
  et~al.}{2001}]{Jones2001}
Jones E.,  Oliphant T.,  Pearu P.,   {Others} 2001, {SciPy: Open Source
  Scientific Tools for Python}, \url {http://www.scipy.org/}

\bibitem[\protect\citeauthoryear{Kant, Puri, Yakovenko  \& Catanzaro}{Kant
  et~al.}{2018}]{Kant2018}
Kant N.,  Puri R.,  Yakovenko N.,   Catanzaro B.,  2018, {Practical Text
  Classification With Large Pre-Trained Language Models}, \url
  {http://arxiv.org/abs/1812.01207}

\bibitem[\protect\citeauthoryear{Kaplan et~al.,}{Kaplan
  et~al.}{2020}]{Kaplan2020}
Kaplan J.,  et~al., 2020, {Scaling laws for neural language models}, \url
  {http://arxiv.org/abs/2001.08361}

\bibitem[\protect\citeauthoryear{Khalifa, Hamed~Taha, Hassanien  \&
  Selim}{Khalifa et~al.}{2018}]{Khalifa2018}
Khalifa N.~E.,  Hamed~Taha M.,  Hassanien A.~E.,   Selim I.,  2018, in 2018
  International Conference on Computing Sciences and Engineering, ICCSE 2018 -
  Proceedings. pp~1--6, \mn@doi{10.1109/ICCSE1.2018.8374210}, \url
  {https://ieeexplore.ieee.org/document/8374210/}

\bibitem[\protect\citeauthoryear{Khramtsov, Dobrycheva, Vasylenko  \&
  Akhmetov}{Khramtsov et~al.}{2019}]{Khrmatsov2019}
Khramtsov V.,  Dobrycheva D.~V.,  Vasylenko M.~Y.,   Akhmetov V.~S.,  2019,
  \mn@doi [Odessa Astronomical Publications]
  {10.18524/1810-4215.2019.32.182092}, 32, 21

\bibitem[\protect\citeauthoryear{Kong, Chen, Chen, Bhatia  \& Callot}{Kong
  et~al.}{2020}]{Kong2020}
Kong L.,  Chen L.,  Chen M.,  Bhatia P.,   Callot L.,  2020, {Improve black-box
  sequential anomaly detector relevancy with limited user feedback}, \url
  {http://arxiv.org/abs/2009.07241}

\bibitem[\protect\citeauthoryear{{LSST Science Collaboration} et~al.,}{{LSST
  Science Collaboration} et~al.}{2009}]{LSSTScienceCollaboration2009}
{LSST Science Collaboration} et~al., 2009, Technical report, {LSST Science
  Book, Version 2.0}, \url {http://arxiv.org/abs/0912.0201}.
LSST Science Collaborations, \url {http://arxiv.org/abs/0912.0201}

\bibitem[\protect\citeauthoryear{Laureijs et~al.,}{Laureijs
  et~al.}{2011}]{Laureijs2011}
Laureijs R.,  et~al., 2011, {Euclid Definition Study Report},
  \mn@doi{10.1088/0264-9381/18/14/306}, \url {http://arxiv.org/abs/1110.3193}

\bibitem[\protect\citeauthoryear{Lavery, Remijan, Charmandaris, Hayes  \&
  Ring}{Lavery et~al.}{2004}]{Lavery2004}
Lavery R.~J.,  Remijan A.,  Charmandaris V.,  Hayes R.~D.,   Ring A.~A.,  2004,
  \mn@doi [The Astrophysical Journal] {10.1086/422420}, 612, 679

\bibitem[\protect\citeauthoryear{LeCun, Bengio  \& Hinton}{LeCun
  et~al.}{2015}]{LeCun2015}
LeCun Y.~A.,  Bengio Y.,   Hinton G.~E.,  2015, \mn@doi [Nature]
  {10.1038/nature14539}, 521, 436

\bibitem[\protect\citeauthoryear{Lintott}{Lintott}{2019}]{Lintott2019Book}
Lintott C.~J.,  2019, {The Crowd and the Cosmos: Adventures in the Zooniverse
  }.
Oxford University Press, Oxford

\bibitem[\protect\citeauthoryear{Lintott et~al.,}{Lintott
  et~al.}{2009}]{Lintott2009}
Lintott C.~J.,  et~al., 2009, \mn@doi [Monthly Notices of the Royal
  Astronomical Society] {10.1111/j.1365-2966.2009.15299.x}, 399, 129

\bibitem[\protect\citeauthoryear{Liu, Ting  \& Zhou}{Liu
  et~al.}{2008}]{Liu2008}
Liu F.~T.,  Ting K.~M.,   Zhou Z.~H.,  2008, in Proceedings - IEEE
  International Conference on Data Mining, ICDM. pp 413--422,
  \mn@doi{10.1109/ICDM.2008.17}

\bibitem[\protect\citeauthoryear{Lochner \& Bassett}{Lochner \&
  Bassett}{2021}]{Lochner2021}
Lochner M.,  Bassett B.~A.,  2021, Astronomy and Computing, 36

\bibitem[\protect\citeauthoryear{Marmanis, Datcu, Esch  \& Stilla}{Marmanis
  et~al.}{2016}]{Marmanis2016}
Marmanis D.,  Datcu M.,  Esch T.,   Stilla U.,  2016, \mn@doi [IEEE Geoscience
  and Remote Sensing Letters] {10.1109/LGRS.2015.2499239}, 13, 105

\bibitem[\protect\citeauthoryear{Martinazzo, Espadoto  \& Hirata}{Martinazzo
  et~al.}{2020}]{Martinazzo2020}
Martinazzo A.,  Espadoto M.,   Hirata N.~S.,  2020, \mn@doi [VISIGRAPP 2020 -
  Proceedings of the 15th International Joint Conference on Computer Vision,
  Imaging and Computer Graphics Theory and Applications]
  {10.5220/0008939800870095}, 5, 87

\bibitem[\protect\citeauthoryear{Mathis, Biasi, Rogers, Bethge  \&
  Weygandt~Mathis}{Mathis et~al.}{2020}]{Mathis2020}
Mathis A.,  Biasi T.,  Rogers B.,  Bethge M.,   Weygandt~Mathis M.,  2020, in
  Uncertainty in Deep Learning.

\bibitem[\protect\citeauthoryear{McInnes, Healy  \& Melville}{McInnes
  et~al.}{2018}]{McInnes2018}
McInnes L.,  Healy J.,   Melville J.,  2018, Journal of Open Source Software,
  3, 861

\bibitem[\protect\citeauthoryear{McKinney}{McKinney}{2010}]{McKinney2010}
McKinney W.,  2010, {Data Structures for Statistical Computing in Python}, \url
  {http://conference.scipy.org/proceedings/scipy2010/mckinney.html}

\bibitem[\protect\citeauthoryear{Mockus \& Mockus}{Mockus \&
  Mockus}{1991}]{Mockus1991}
Mockus J.~B.,  Mockus L.~J.,  1991, \mn@doi [Journal of Optimization Theory and
  Applications] {10.1007/BF00940509}, 70, 157

\bibitem[\protect\citeauthoryear{Moiseev, Smirnova, Smirnova  \&
  Reshetnikov}{Moiseev et~al.}{2011}]{Moiseev2011}
Moiseev A.~V.,  Smirnova K.~I.,  Smirnova A.~A.,   Reshetnikov V.~P.,  2011,
  \mn@doi [Monthly Notices of the Royal Astronomical Society]
  {10.1111/j.1365-2966.2011.19479.x}, 418, 244

\bibitem[\protect\citeauthoryear{Moosavi-Dezfooli, Fawzi, Fawzi  \&
  Frossard}{Moosavi-Dezfooli et~al.}{2017}]{Moosavi-Dezfooli2017}
Moosavi-Dezfooli S.~M.,  Fawzi A.,  Fawzi O.,   Frossard P.,  2017, in
  Proceedings - 30th IEEE Conference on Computer Vision and Pattern
  Recognition, CVPR 2017. Institute of Electrical and Electronics Engineers
  Inc., pp 86--94, \mn@doi{10.1109/CVPR.2017.17}, \url
  {http://arxiv.org/abs/1610.08401}

\bibitem[\protect\citeauthoryear{Murphy}{Murphy}{2012}]{Murphy2012}
Murphy K.~P.,  2012, {Machine Learning: A Probabilisitic Perspective}.
MIT Press, Boston, MA

\bibitem[\protect\citeauthoryear{Nair \& Abraham}{Nair \&
  Abraham}{2010}]{Nair2010}
Nair P.~B.,  Abraham R.~G.,  2010, \mn@doi [The Astrophysical Journal
  Supplement Series] {10.1088/0067-0049/186/2/427}, 186, 427

\bibitem[\protect\citeauthoryear{Pedregosa et~al.,}{Pedregosa
  et~al.}{2012}]{Pedregosa2012}
Pedregosa F.,  et~al., 2012, \mn@doi [Journal of Machine Learning Research]
  {10.1007/s13398-014-0173-7.2}, 12, 2825

\bibitem[\protect\citeauthoryear{Pelleg, Pelleg  \& Moore}{Pelleg
  et~al.}{2004}]{Pelleg2014}
Pelleg D.,  Pelleg D.,   Moore A.,  2004, in Proceedings of the 17th
  International Conference on Neural Information Processing Systems. No.
  January 2004.
Vancouver, Canada, pp 1073--1080, \url
  {https://dl.acm.org/doi/10.5555/2976040.2976175}

\bibitem[\protect\citeauthoryear{P{\'{e}}rez-Carrasco, Cabrera-Vives,
  Martinez-Marin, Cerulo, Demarco, Protopapas, Godoy  \&
  Huertas-Company}{P{\'{e}}rez-Carrasco et~al.}{2019}]{Perez-Carrasco2018}
P{\'{e}}rez-Carrasco M.,  Cabrera-Vives G.,  Martinez-Marin M.,  Cerulo P.,
  Demarco R.,  Protopapas P.,  Godoy J.,   Huertas-Company M.,  2019, \mn@doi
  [Publications of the Astronomical Society of the Pacific]
  {10.1088/1538-3873/aaeeb4}, 131

\bibitem[\protect\citeauthoryear{Ralph et~al.,}{Ralph et~al.}{2019}]{Ralph2019}
Ralph N.~O.,  et~al., 2019, \mn@doi [Publications of the Astronomical Society
  of the Pacific] {10.1088/1538-3873/ab213d}, 131, 108011

\bibitem[\protect\citeauthoryear{Rasmussen \& Williams}{Rasmussen \&
  Williams}{2006}]{Rasmussen2006}
Rasmussen C.~E.,  Williams C. K.~I.,  2006, {Gaussian Processes for Machine
  Learning}.
Adaptive Computation and Machine Learning, MIT Press, Cambridge, MA, USA

\bibitem[\protect\citeauthoryear{Recht, Roelofs, Schmidt  \& Shankar}{Recht
  et~al.}{2019}]{Recht2019}
Recht B.,  Roelofs R.,  Schmidt L.,   Shankar V.,  2019, in 36th International
  Conference on Machine Learning, ICML 2019. International Machine Learning
  Society (IMLS), pp 9413--9424, \url {http://arxiv.org/abs/1902.10811}

\bibitem[\protect\citeauthoryear{Reis, Rotman, Poznanski, Prochaska  \&
  Wolf}{Reis et~al.}{2021}]{Reis2019}
Reis I.,  Rotman M.,  Poznanski D.,  Prochaska J.~X.,   Wolf L.,  2021, \mn@doi
  [Astronomy and Computing] {10.1016/j.ascom.2020.100437}, 34, 100437

\bibitem[\protect\citeauthoryear{Ridnik, Ben-Baruch, Noy  \&
  Zelnik-Manor}{Ridnik et~al.}{2021}]{Ridnik2021}
Ridnik T.,  Ben-Baruch E.,  Noy A.,   Zelnik-Manor L.,  2021, in Proceedings of
  Neural Information Processing Systems. \url {http://arxiv.org/abs/2104.10972}

\bibitem[\protect\citeauthoryear{Ross, Lim, Lin  \& Yang}{Ross
  et~al.}{2008}]{Ross2008}
Ross D.~A.,  Lim J.,  Lin R.~S.,   Yang M.~H.,  2008, \mn@doi [International
  Journal of Computer Vision] {10.1007/s11263-007-0075-7}, 77, 125

\bibitem[\protect\citeauthoryear{Russakovsky et~al.,}{Russakovsky
  et~al.}{2015}]{Russakovsky2015}
Russakovsky O.,  et~al., 2015, \mn@doi [International Journal of Computer
  Vision] {10.1007/s11263-015-0816-y}, 115, 211

\bibitem[\protect\citeauthoryear{Sandler, Howard, Zhu, Zhmoginov  \&
  Chen}{Sandler et~al.}{2018}]{Sandler2018}
Sandler M.,  Howard A.,  Zhu M.,  Zhmoginov A.,   Chen L.~C.,  2018, in
  Proceedings of the IEEE Computer Society Conference on Computer Vision and
  Pattern Recognition. IEEE Computer Society, pp 4510--4520,
  \mn@doi{10.1109/CVPR.2018.00474}, \url {http://arxiv.org/abs/1801.04381}

\bibitem[\protect\citeauthoryear{Sarmiento, Huertas-Company, Knapen,
  S{\'{a}}nchez, S{\'{a}}nchez, Drory  \& Falc{\'{o}}n-Barroso}{Sarmiento
  et~al.}{2021}]{Sarmiento2021}
Sarmiento R.,  Huertas-Company M.,  Knapen J.~H.,  S{\'{a}}nchez S.~F.,
  S{\'{a}}nchez H.~D.,  Drory N.,   Falc{\'{o}}n-Barroso J.,  2021, {Capturing
  the physics of MaNGA galaxies with self-supervised Machine Learning}, \url
  {http://arxiv.org/abs/2104.08292}

\bibitem[\protect\citeauthoryear{Schutter \& Shamir}{Schutter \&
  Shamir}{2015}]{Schutter2015}
Schutter A.,  Shamir L.,  2015, \mn@doi [Astronomy and Computing]
  {10.1016/j.ascom.2015.05.002}, 12, 60

\bibitem[\protect\citeauthoryear{Schwarz}{Schwarz}{1981}]{Schwarz1981}
Schwarz M.~P.,  1981, \mn@doi [The Astrophysical Journal] {10.1086/159011},
  247, 77

\bibitem[\protect\citeauthoryear{Shamir}{Shamir}{2020}]{Shamir2020Rings}
Shamir L.,  2020, \mn@doi [Monthly Notices of the Royal Astronomical Society]
  {10.1093/mnras/stz3297}, 491, 3767

\bibitem[\protect\citeauthoryear{Sharma \& Kaplan}{Sharma \&
  Kaplan}{2020}]{Sharma2020}
Sharma U.,  Kaplan J.,  2020, {A neural scaling law from the dimension of the
  data manifold}, \url {https://arxiv.org/abs/2004.10802}

\bibitem[\protect\citeauthoryear{Siddiqui, Wright, Fern, Theriault, Dietterich
  \& Archer}{Siddiqui et~al.}{2018}]{Siddiqui2018}
Siddiqui M.~A.,  Wright R.,  Fern A.,  Theriault A.,  Dietterich T.~G.,
  Archer D.~W.,  2018, in Proceedings of the ACM SIGKDD International
  Conference on Knowledge Discovery and Data Mining. Association for Computing
  Machinery, pp 2200--2209, \mn@doi{10.1145/3219819.3220083}

\bibitem[\protect\citeauthoryear{Simmons et~al.,}{Simmons
  et~al.}{2013}]{Simmons2013}
Simmons B.,  et~al., 2013, \mn@doi [Monthly Notices of the Royal Astronomical
  Society] {10.1093/mnras/sts491}, 429, 2199

\bibitem[\protect\citeauthoryear{Simonyan \& Zisserman}{Simonyan \&
  Zisserman}{2015}]{Simonyan2014}
Simonyan K.,  Zisserman A.,  2015, in International Conference on Learning
  Representations. , \mn@doi{10.1016/j.infsof.2008.09.005}, \url
  {http://arxiv.org/abs/1409.1556}

\bibitem[\protect\citeauthoryear{Smethurst, Lintott, Bamford, Hart, Kruk,
  Masters, Nichol  \& Simmons}{Smethurst et~al.}{2017}]{Smethurst2017}
Smethurst R.~J.,  Lintott C.~J.,  Bamford S.~P.,  Hart R.~E.,  Kruk S.~J.,
  Masters K.~L.,  Nichol R.~C.,   Simmons B.~D.,  2017, \mn@doi [Monthly
  Notices of the Royal Astronomical Society] {10.1093/mnras/stx973}, 469, 3670

\bibitem[\protect\citeauthoryear{Spindler, Geach  \& Smith}{Spindler
  et~al.}{2020}]{Spindler2020}
Spindler A.,  Geach J.~E.,   Smith M.~J.,  2020, \mn@doi [Monthly Notices of
  the Royal Astronomical Society] {10.1093/mnras/staa3670}, 502, 985

\bibitem[\protect\citeauthoryear{Stein, Harrington, Blaum, Medan  \&
  Lukic}{Stein et~al.}{2021}]{Stein2021}
Stein G.,  Harrington P.,  Blaum J.,  Medan T.,   Lukic Z.,  2021,
  {Self-supervised similarity search for large scientific datasets}, \url
  {http://arxiv.org/abs/2110.13151}

\bibitem[\protect\citeauthoryear{Storey-Fisher, Huertas-Company, Ramachandra,
  Lanusse, Leauthaud, Luo, Huang  \& Prochaska}{Storey-Fisher
  et~al.}{2021}]{Storey-Fisher2021AnomalyNetworks}
Storey-Fisher K.,  Huertas-Company M.,  Ramachandra N.,  Lanusse F.,  Leauthaud
  A.,  Luo Y.,  Huang S.,   Prochaska J.~X.,  2021, Monthly Notices of the
  Royal Astronomical Society, 508, 2946

\bibitem[\protect\citeauthoryear{Struck}{Struck}{2010}]{Struck2010}
Struck C.,  2010, \mn@doi [Monthly Notices of the Royal Astronomical Society]
  {10.1111/j.1365-2966.2009.16224.x}, 403, 1516

\bibitem[\protect\citeauthoryear{Szegedy et~al.,}{Szegedy
  et~al.}{2015}]{Szegedy2015Deeper}
Szegedy C.,  et~al., 2015, in Proceedings of the IEEE Computer Society
  Conference on Computer Vision and Pattern Recognition. ,
  \mn@doi{10.1109/CVPR.2015.7298594}, \url {http://arxiv.org/abs/1409.4842}

\bibitem[\protect\citeauthoryear{Tan \& Le}{Tan \& Le}{2019}]{Tan2019a}
Tan M.,  Le Q.~V.,  2019, in 36th International Conference on Machine Learning,
  ICML 2019. pp 10691--10700, \url {http://arxiv.org/abs/1905.11946}

\bibitem[\protect\citeauthoryear{Tang, Scaife  \& Leahy}{Tang
  et~al.}{2019}]{Tang2019}
Tang H.,  Scaife A.~M.,   Leahy J.~P.,  2019, \mn@doi [Monthly Notices of the
  Royal Astronomical Society] {10.1093/mnras/stz1883}, 488, 3358

\bibitem[\protect\citeauthoryear{{The Astropy Collaboration} et~al.,}{{The
  Astropy Collaboration} et~al.}{2018}]{TheAstropyCollaboration2018}
{The Astropy Collaboration} et~al., 2018, \mn@doi [The Astronomical Journal]
  {arXiv:1801.02634v2}, 156, 123

\bibitem[\protect\citeauthoryear{Timmis \& Shamir}{Timmis \&
  Shamir}{2017}]{Timmis2017}
Timmis I.,  Shamir L.,  2017, \mn@doi [The Astrophysical Journal Supplement
  Series] {10.3847/1538-4365/aa78a3}, 231, 2

\bibitem[\protect\citeauthoryear{Tschandl, Sinz  \& Kittler}{Tschandl
  et~al.}{2019}]{Tschandl2019}
Tschandl P.,  Sinz C.,   Kittler H.,  2019, \mn@doi [Computers in Biology and
  Medicine] {10.1016/j.compbiomed.2018.11.010}, 104, 111

\bibitem[\protect\citeauthoryear{Van Den~Oord, Kalchbrenner, Vinyals, Espeholt,
  Graves  \& Kavukcuoglu}{Van Den~Oord et~al.}{2016}]{VanDenOord2016a}
Van Den~Oord A.,  Kalchbrenner N.,  Vinyals O.,  Espeholt L.,  Graves A.,
  Kavukcuoglu K.,  2016, in Advances in Neural Information Processing Systems.
  Neural information processing systems foundation, pp 4797--4805, \url
  {http://arxiv.org/abs/1606.05328}

\bibitem[\protect\citeauthoryear{Variawa, van Zyl  \& Woolway}{Variawa
  et~al.}{2020}]{Variawa2020}
Variawa M.~Z.,  van Zyl T.~L.,   Woolway M.,  2020, \mn@doi [2020 IEEE 23rd
  International Conference on Information Fusion (FUSION)]
  {10.23919/FUSION45008.2020.9190462}, pp~1--7

\bibitem[\protect\citeauthoryear{Vaswani, Shazeer, Parmar, Uszkoreit, Jones,
  Gomez, Kaiser  \& Polosukhin}{Vaswani et~al.}{2017}]{Vaswani2017}
Vaswani A.,  Shazeer N.,  Parmar N.,  Uszkoreit J.,  Jones L.,  Gomez A.~N.,
  Kaiser L.,   Polosukhin I.,  2017, Advances in Neural Information Processing
  Systems, 2017-Decem, 5999

\bibitem[\protect\citeauthoryear{Walmsley et~al.,}{Walmsley
  et~al.}{2020}]{Walmsley2020}
Walmsley M.,  et~al., 2020, \mn@doi [Monthly Notices of the Royal Astronomical
  Society] {10.1093/mnras/stz2816}, 491, 1554

\bibitem[\protect\citeauthoryear{Walmsley et~al.,}{Walmsley
  et~al.}{2022}]{Walmsley2022decals}
Walmsley M.,  et~al., 2022, Monthly Notices of the Royal Astronomical Society,
  509, 3966

\bibitem[\protect\citeauthoryear{Welsh et~al.,}{Welsh et~al.}{2011}]{Welsh2011}
Welsh W.~F.,  et~al., 2011, \mn@doi [Astrophysical Journal, Supplement Series]
  {10.1088/0067-0049/197/1/4}, 197, 4

\bibitem[\protect\citeauthoryear{Willett et~al.,}{Willett
  et~al.}{2013}]{Willett2013}
Willett K.~W.,  et~al., 2013, \mn@doi [Monthly Notices of the Royal
  Astronomical Society] {10.1093/mnras/stt1458}, 435, 2835

\bibitem[\protect\citeauthoryear{Wu et~al.,}{Wu et~al.}{2018}]{Wu2018}
Wu C.,  et~al., 2018, \mn@doi [Monthly Notices of the Royal Astronomical
  Society] {10.1093/mnras/sty2646}, 1230, 1211

\bibitem[\protect\citeauthoryear{Yang, Qinami, Fei-Fei, Deng  \&
  Russakovsky}{Yang et~al.}{2020a}]{YangFair2020}
Yang K.,  Qinami K.,  Fei-Fei L.,  Deng J.,   Russakovsky O.,  2020a, in FAT*
  2020 - Proceedings of the 2020 Conference on Fairness, Accountability, and
  Transparency. Association for Computing Machinery, Inc, pp 547--558,
  \mn@doi{10.1145/3351095.3375709}

\bibitem[\protect\citeauthoryear{Yang, Zhu, Gmyr, Zeng, Huang  \& Darve}{Yang
  et~al.}{2020b}]{YangZ2020}
Yang Z.,  Zhu C.,  Gmyr R.,  Zeng M.,  Huang X.,   Darve E.,  2020b, in EMNLP
  2020. Association for Computational Linguistics (ACL), pp 1865--1874, \url
  {http://arxiv.org/abs/2001.00725}

\bibitem[\protect\citeauthoryear{Zanisi et~al.,}{Zanisi
  et~al.}{2021}]{Zanisi2021}
Zanisi L.,  et~al., 2021, \mn@doi [Monthly Notices of the Royal Astronomical
  Society] {10.1093/mnras/staa3864}, 501, 4359

\bibitem[\protect\citeauthoryear{van~der Walt, Sch{\"{o}}nberger,
  Nunez-Iglesias, Boulogne, Warner, Yager, Gouillart  \& Yu}{van~der Walt
  et~al.}{2014}]{VanderWalt2014}
van~der Walt S.,  Sch{\"{o}}nberger J.~L.,  Nunez-Iglesias J.,  Boulogne F.,
  Warner J.~D.,  Yager N.,  Gouillart E.,   Yu T.,  2014, \mn@doi [PeerJ]
  {10.7717/peerj.453}, 2, e453

\makeatother
\end{thebibliography}

\appendix

\section{Further Visualisations}
\label{sec:further_viz}

\begin{figure*}
    \centering
    \includegraphics[width=\textwidth]{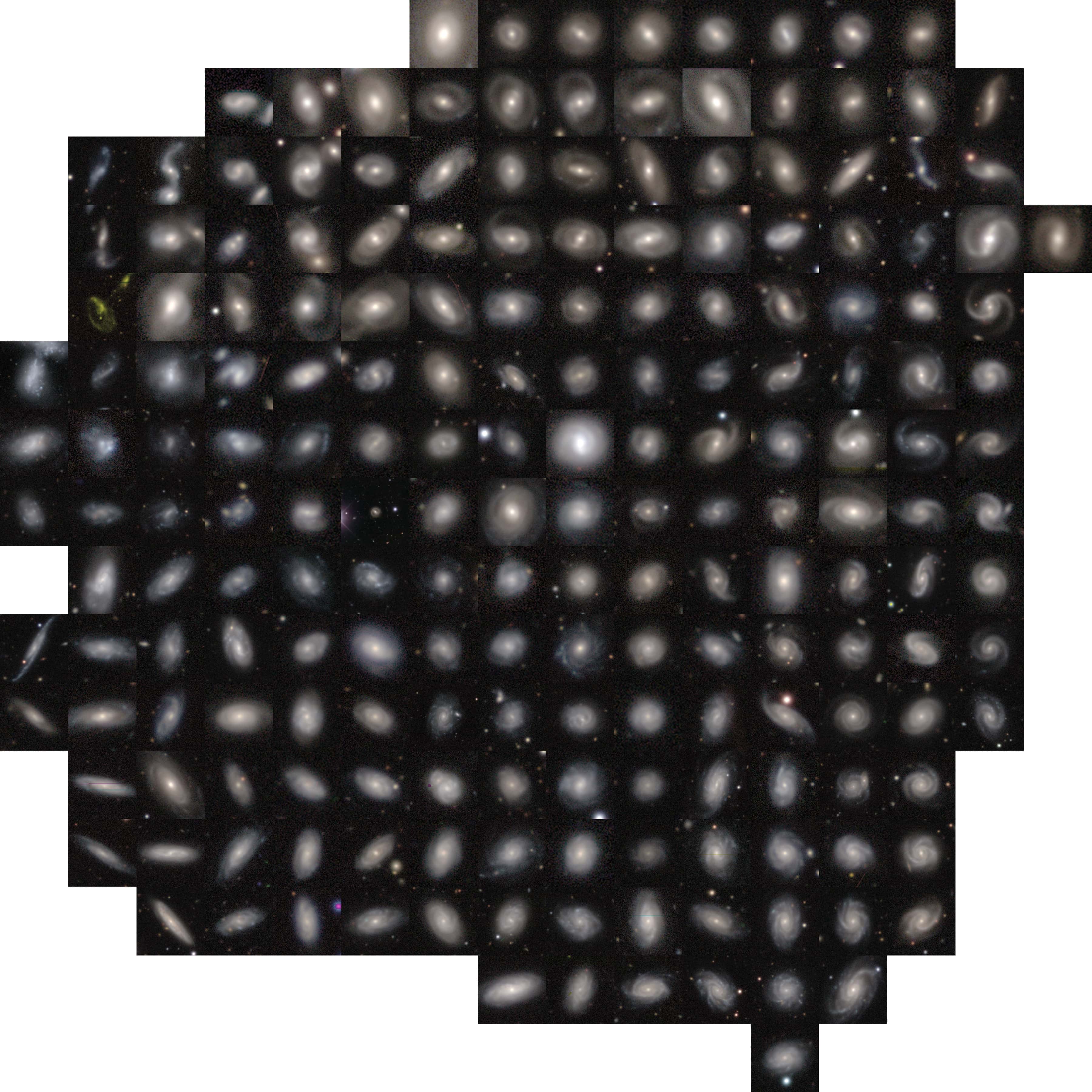}
    \caption{As with Fig. \ref{fig:umap_viz_all}, a visualisation of the representation learned by our CNN, showing similar galaxies occupying similar regions of feature space. Created using Incremental PCA and \texttt{umap} to compress the representation to 2D, and then placing galaxy thumbnails at the 2D location of the corresponding galaxy. Galaxies are filtered to be featured and face-on (specifically, $f_\text{feat} \times f_\text{face} > 0.5$, where $f$ is the GZ DECaLS automatic vote fraction for each answer).}
    \label{fig:umap_viz_feat}
\end{figure*}

\begin{figure*}
    \centering
    \includegraphics[width=\textwidth]{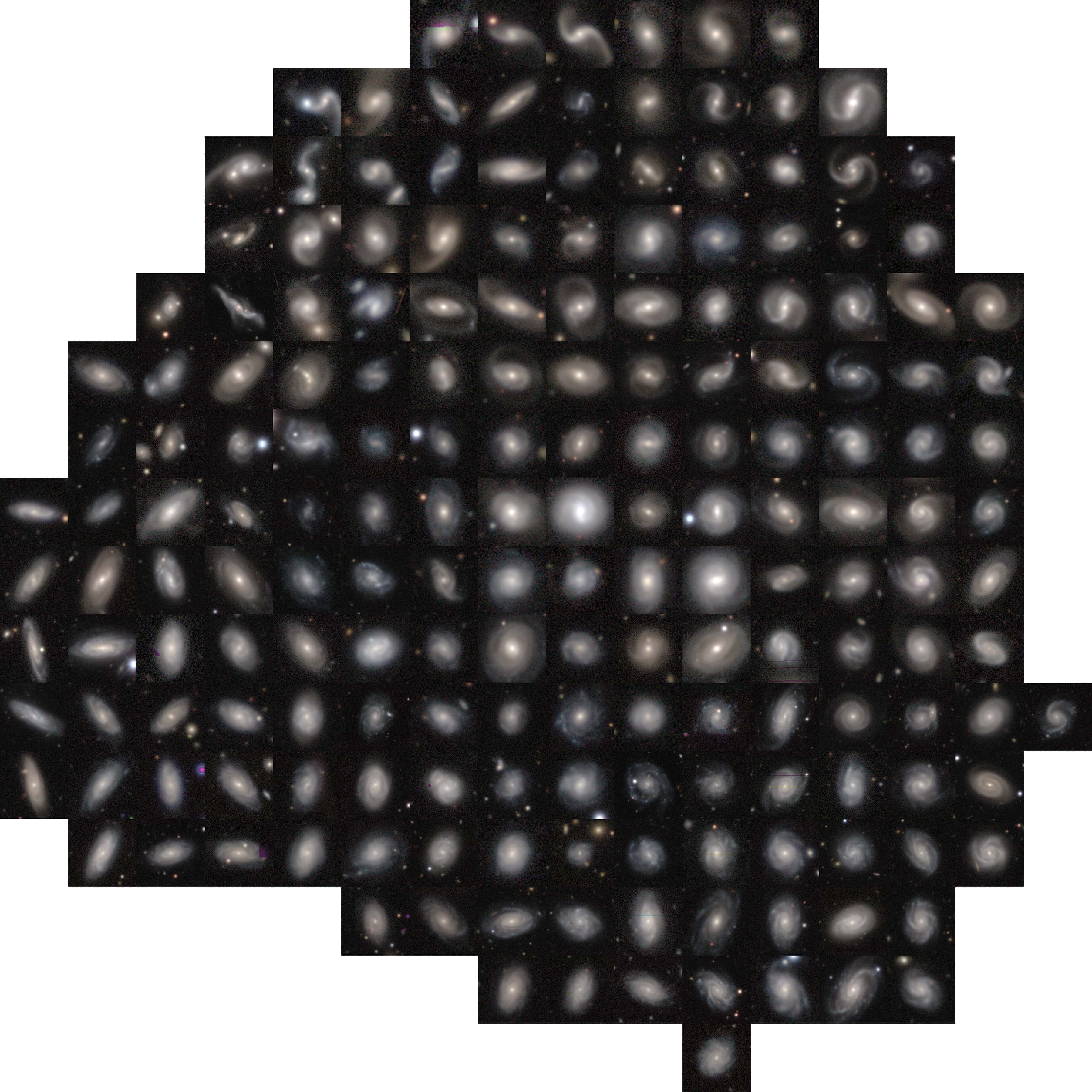}
    \caption{As with Fig. \ref{fig:umap_viz_all} and Fig. \ref{fig:umap_viz_feat}, but filtering galaxies to be spirals (specifically, $f_\text{feat} \times f_\text{face} \times f_\text{spiral} > 0.5$, where $f$ is the GZ DECaLS automatic vote fraction for each answer).}
    \label{fig:umap_viz_spiral}
\end{figure*}


\bsp	
\label{lastpage}
\end{document}